\documentclass{aa}  

\usepackage{graphicx}
\usepackage{subfig}
\usepackage{appendix}
\usepackage{verbatim}
\usepackage{txfonts}
\usepackage[switch]{lineno}
\usepackage{xcolor}
\usepackage{ulem}
\usepackage{longtable}
\usepackage{natbib}
\usepackage{float}
\bibpunct{(}{)}{;}{a}{}{,} 

\begin{document} 

   \title{Exploring short-term stellar activity in M dwarfs: A volume-limited perspective}


   \author{G. Galletta$^{1,2}$, S. Colombo$^3$, L. Prisinzano$^3$ and G. Micela$^3$ }

   \institute{$^1$Dipartimento di Fisica e  Chimica, Università di Palermo, Via Archirafi 36, 90128, Palermo, Italy.\\
    $^2$Blue Skies Space Italia S.R.L.
    Via Vincenzo Monti 16, 20123, Milano, Italy.\\
    $^3$INAF, Osservatorio Astronomico di
    Palermo, Palermo, Italy.\\
    \email{gabriele.galletta@inaf.it}
    }

    \date{Received ; accepted }

 
  \abstract
   {
   Flares are a form of stellar activity that occur over short timescales but produce highly energetic outbursts. Studying stellar flares is crucial because they can significantly alter the circumstellar environment by producing intense high-energy radiation.
   Understanding stellar flares is essential for clarifying the environment in which planets evolve, as flares can influence planetary atmospheres by driving photoevaporation and photochemical processes. M dwarfs are of significant interest due to their high flare activity rates and the potential presence of exoplanets within their habitable zones, whose atmospheres may be influenced by flare-emitted radiation.
   }
   {We aimed to define the flaring properties of an unbiased sample of M dwarfs with limited volume. Using data from the Transiting Exoplanet Survey Satellite (TESS), we characterized the frequency, energy distribution, and temporal properties of flares in nearby stars.}
   {We selected a volume-limited sample of M dwarfs within 10 pc from Earth from the Gaia DR3 catalog. We analyzed TESS light curves using an iterative Gaussian process fitting technique to remove long-term stellar activity signals, enabling the identification and characterization of impulsive flare events. For each flare, we derived the amplitudes, timescales, and total energy emitted.}
   {We analyzed 173 stars and detected 17,229 flares, with 0 to 76 flares per TESS sector. We examined the frequency and energy distribution of stellar flares using three representative stars to illustrate the diversity in flare activity. We observed flares with a minimum energy of $\sim 10^{29}$ erg and typical durations ranging from 2 to 8000 seconds. We modeled the cumulative flare energy distribution using one-slope and two-slope fits, yielding average slopes of -0.79$\pm$0.64 and -1.23$\pm$1.32, respectively.
   We defined the Flare Energy Index (GF.01) to characterize the flare frequency and revealed two distinct populations.  Fainter stars exhibited fewer high-energy flares, whereas brighter stars exhibited more frequent low-energy flares. We analyzed two highly active stars with the largest number of TESS sectors, G 227-22 and G 258-33, were analyzed over a long time baseline to explore their flare properties and energy distributions.}
   {}


   \maketitle
%

\section{Introduction}
The brightness of solar-type and smaller stars varies over time due to a variety of events collectively referred to as stellar activity \citep{radick1990stellar}. These phenomena result from the stellar dynamo \citep{donati2008magnetic}, stellar cycles \citep{strugarek2014diversity}, magnetic reconnection events \citep{lanza2009stellar}, and, in some cases, interactions between stars and planets \citep{cauley2019magnetic}. 
Flares are sudden, powerful releases of energy from stars, occuring over short timescales and involving intense bursts of radiation. Studying flares enhances our understanding of magnetic field generation and evolution through stellar dynamo processes and magnetic reconnection events, as first described for solar flares by \cite{carmichael196454}. Moreover, understanding the occurrence of flares sheds light on the parameters that influence planetary atmospheres and their chemical compositions \citep{ buccino2007uv, segura2010effect,johnstone2016influences}.\\
M dwarfs are of particular interest due to their high flare activity rates \citep{hilton2011galactic} and the possibility of hosting exoplanets within their habitable zones. These habitable zones lie close to the star, making planets especially susceptible to the effects of flares \citep{bonfils2013harps}. \\ 
The Transiting Exoplanet Survey Satellite \citep[TESS,][]{ricker2016transiting} mission offers a powerful platform to observe star activity over vast regions of the sky. Created to detect exoplanets using the transit method, its wide field of view and high temporal cadence make TESS particularly useful for observing rapid brightness variations in stars, such as flares.\\
In this study, we take advantage of TESS's capabilities to define the flaring properties of an unbiased volume-limited sample of M dwarfs within 10 parsecs of Earth. We focus on determining the frequency, energy distribution, and temporal properties of flares in these nearby stars. 
Previous studies have investigated the energy distributions of flares across various wavelength ranges. For example, \cite{haisch1983x} examined X-ray emissions from stellar flares and found that these emissions vary in intensity and energy depending on the type of star, providing insights into their energy distributions. \cite{haisch1991flares} discusses the characteristics of stellar flares, including X-ray and UV emissions, comparing the Sun's activity with that of other stars. While the emission mechanisms are similar, there are significant differences in energy output and duration across different types of stars. 
\cite{mitra2005relationship} observed a correlation between the X-ray and ultraviolet fluxes for stellar flares, indicating a power relationship between these two wavelengths.
\cite{feinstein2024evolution} analyzed young stars (<300 Myr) in 26 nearby moving groups, associations, or clusters, monitored by TESS at a 2-minute cadence. Their study identified rotation periods for 1,847 stars and detected 26,355 flares from 3,160 stars. While that work provides a foundation for understanding short-term variability in young stars, our study focuses on an older and closer population of stars. In a complementary study, \cite{stelzer2022flares} analyzed a sample of stars from the TESS Habitable Zone Star Catalog \citep[HZCat][]{kaltenegger2019tess}, a subset of the TESS Input Catalog (TIC). This list includes 1,822 stars with TESS magnitudes below 12, allowing the detection of planets up to two Earth radii in size. While their study focuses on stars with planets in the habitable zone, we investigate the general properties of stellar variability in M dwarfs.\\ 
Gaia Data Release 3 (DR3) includes processing and analysis of photometric and spectroscopic variability for 1.8 billion sources, providing valuable data for the study of stellar activity. \cite{apellaniz2023stellar} analyzed rotational modulation and color variation patterns in Sun-like stars, providing insights into stellar magnetic activity. Their study, based on Gaia DR3 data, revealed periodic brightness variations caused by starspots and rotational modulation, as well as color changes linked to magnetic cycles. This work provides a large-scale perspective on how stellar magnetism evolves over time.
Several studies have leveraged Gaia data to investigate stellar variability and magnetic activity, providing information into the behavior of low-mass stars. \cite{distefano2023gaia} analyzed rotational modulation and chromatic variations in Sun-like stars, revealing a correlation between magnetic activity and photometric variability. 
Building on Gaia's capabilities, \cite{lanzafame2018gaia} developed a method to identify low-mass variable stars and characterize their variability parameters. Their analysis provides a quantitative framework for describing the distribution of such stars in the Hertzsprung-Russell (HR) diagram.
In a complementary study, \cite{lanzafame2023gaia} derived an activity index from the Ca II triplet using Gaia spectra, offering a new diagnostic tool for assessing magnetic activity in low-mass stars.\\
Our study aims to shed light on the characteristics of short-term stellar variability in M dwarfs using a volume-limited sample that represents the bulk properties of M dwarfs in the solar neighborhood in an unbiased manner. M dwarfs are the most abundant stellar type in the Galaxy and are key targets in the search for habitable exoplanets due to their small size and low luminosity, making planetary signals easier to detect and study \citep{shields2016habitability}. Understanding their variability is crucial for characterizing their magnetic activity and assessing its impact on planetary atmospheres \citep{wunderlich2019detectability}. In this article, we analyze all M dwarfs in the Gaia DR3 catalog \citep{prusti2016gaia, vallenari2023gaia} within 10 parsecs that have been observed by TESS through sector 76. We implement a new automated procedure to fit cumulative flare energy distributions to determine the flare frequency for each star.\\
Building on the methodology developed by \cite{colombo2022short}, this study adapts their flare-fitting tool, originally applied to the stars DS Tuc and AU Mic, to analyze a much larger sample of stars using TESS light curves.\\
The paper is structured as follows. Section \ref{method} describes the sample, the method used to analyze flare activity, the calculation of bolometric luminosity and the automated procedure to calculate the slopes and energy breaks of the cumulative distribution. Sect. \ref{results} shows the results obtained from the analysis. In Sect. \ref{discussion} we discuss our results and Sect. \ref{conclusions} summarizes our conclusions.

\section{Methods}\label{method}
In this section, we describe the sample and the method used to analyze it. In Sect. \ref{sample} we describe the sample, and
in Subsect. \ref{bolometricluminosity} we detail how bolometric luminosities are calculated.
In Sect. \ref{flarefitting} we discuss how flares are identified and fitted. 
In Sect. \ref{cumulative} we present the cumulative frequency energy distribution method.
In Sect. \ref{slopecalculations} we present the slope and energy breaks calculation method.

\subsection{The sample}\label{sample}
\begin{figure}
    \centering
    \includegraphics[width=\hsize]{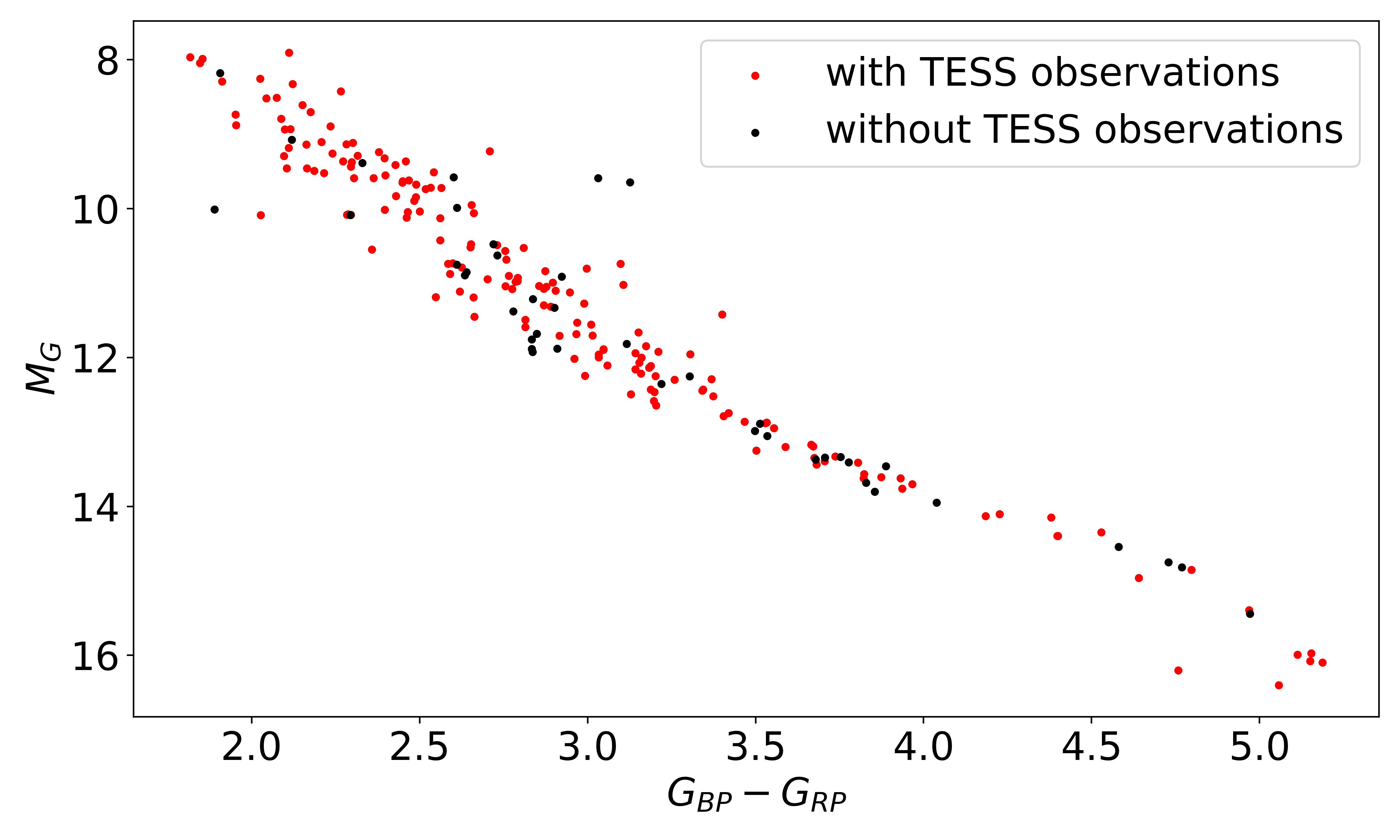}
    \caption{Color-absolute magnitude diagram of the sample. Sample without TESS observations are shown in black, while sample with TESS observations are shown in red.}
    \label{fig:hrdiagram}
\end{figure}

We selected our sample from the Gaia ESA archive using the astronomical data query language (ADQL). To select M dwarfs within 10 parsecs of the Sun, we applied the following criteria based on Gaia DR3 measurements: the parallax ($\bar{\omega}$), expressed in milliarcseconds, had to be $\bar{\omega} > 100$ mas; the Gaia blue and red photometric magnitudes, $G_{BP}$ and $G_{RP}$ respectively, had to satisfy $G_{BP} - G_{RP} > 1.8$; the apparent G-band magnitude ($G$) had to be $G < 16$; and the relative parallax uncertainty, defined as $\sigma_{\bar{\omega}} / \bar{\omega}$, had to be $< 0.2$.\\
The first condition ensures that only stars within <10 pc were selected. We considered parallax uncertainty lower than 20\%, ensuring reliable distance estimates. The second condition selected only M dwarfs, which according to the color-temperature table of \cite{pecaut2013intrinsic}, have $G_{BP}$-$G_{RP}$> 1.8.\footnote{\url{https://www.pas.rochester.edu/~emamajek/EEM_dwarf_UBVIJHK_colors_Teff.txt}} We note that at distances less than 10 pc, the reddening is negligible and can be ignored.
The third criterion ($G$ < 16) was used to select stars with magnitudes within the TESS magnitude limit.
Applying these selection criteria, we retrieved 213 stars, of which 173 have counterparts in the TESS catalog.\\
In Fig. \ref{fig:hrdiagram}, we present the color-absolute magnitude diagram of the sample. Notably, for values of $M_G > 12.58$, the data spread narrows, indicating that stars in this magnitude range are more tightly clustered compared to lower-magnitude stars. From this point on, we consider $M_G=12.58$ as the magnitude threshold to distinguish between the two subsamples.
The observed spread in the color-absolute magnitude diagram for lower magnitudes is consistent with typical patterns seen in Gaia data and is likely attributed to stellar variability. \cite{eyer2019gaia} illustrate (in their Fig. 8) that the fraction of variable stars decreases significantly for $G_{BP}$-$G_{RP}$ values approximately greater than 3.3, which corresponds to the threshold where we began to observe a narrow sequence.
However, stellar activity is expected to produce photometric variability and the phenomenon known as radius inflation, which might contribute to the spread observed in the color-magnitude diagram of M dwarfs. Radius inflation refers to an increase in the stellar radius compared to theoretical predictions, which can affect the star's brightness. Recent studies, such as \cite{wanderley2023stellar}, show that radius inflation is more pronounced in partially convective M dwarfs and rapid rotators, with an inflation of up to 2.7\%$\pm$2.1\%, potentially linked to magnetic fields. \\Similarly, \cite{kiman2024accurate} found a correlation between magnetic activity, measured via $H\alpha$ emission, and radius inflation in field stars. Active stars tend to have larger radii than inactive stars, with a maximum inflation of 9\%. This effect is most noticeable for stars with masses between 0.5 and 0.6 $M_\odot$, while no significant correlation was found for lower-mass stars, possibly due to measurement uncertainties or physical reasons. These results suggest that radius inflation may contribute to the observed spread in the color-magnitude diagram, particularly among early M dwarfs.\\
Using Gaia DR3 identifiers, we retrieved the corresponding TIC identifiers through the SIMBAD astronomical database\footnote{http://simbad.u-strasbg.fr/simbad/} \citep{wenger2000simbad}. We then employed the Lightkurve package \citep{2018ascl.soft12013L} to retrieve the TESS data by specifying search parameters such as an exposure time of 120 seconds and a search radius of 0.5 arcsec. Specifically, we searched for light curves with short-cadence data from the SPOC (Science Processing Operations Center) pipeline. The target columns retrieved for analysis were \textit{TIME} and \textit{SAP\_FLUX}.\\
The sample showed a distribution of stars with an average availability of four TESS sectors, with only a few stars having data from 24 or more sectors. In total, we analyzed 751 TESS sectors. Table 3 lists the names of the stars analyzed, the TESS sectors, the number of flares, the $L_{bol}$, the slopes of the cumulative energy distributions, and the energy breaks.

\subsection{Calculation of bolometric luminosity}\label{bolometricluminosity}
To calculate the energy released by each flare following the method of \cite{colombo2022short}, we require the bolometric luminosity of all stars in the sample.\\  \cite{bonfils2013harps} estimated the bolometric luminosity by utilizing photometric data and empirical relations. They specifically employed bolometric corrections derived from empirical calibrations linking bolometric magnitudes to optical and infrared magnitudes. This method adjusts measured magnitudes to account for the star's spectral energy distribution (SED), thereby compensating for the star's overall energy output.  Similarly, \cite{ribas2017full} and \cite{paudel2021simultaneous} computed bolometric luminosities by fitting stellar parameters from SEDs constructed by combining distance measurements with flux data at various wavelengths. The bolometric luminosity was then calculated by integrating the total luminosity throughout the entire SED. This approach covers a broad range of wavelengths, providing an accurate measure of the star's overall energy output.\\
In this work, we used the Virtual Observatory SED Analyzer (VOSA) to compute the bolometric luminosity \citep{rodrigo2020vosa}. VOSA constructs and analyses SEDs based on observational flux data, distance in parsecs, and sky locations (RA and DEC). To calculate the bolometric luminosity, we integrated the star's emitted energy across the entire range of wavelengths by combining photometric data from various catalogs, thereby extending the available wavelength coverage from the UV to the IR regions. The tool fits the observational data using theoretical models of the stellar atmosphere and determines the brightness and other physical parameters of the stars. To fit the observational data, we employed the theoretical spectra model grid of BT-Settl (CIFIST)\citep{allard2013bt} which incorporates a cloud model valid over the entire parameter range and uses solar abundances from \cite{caffau2011solar}.
To validate our method, Table \ref{tab:tablbol} compares the estimated bolometric luminosities of HD 197481, AD Leo, EV Lac, Proxima Centauri, Wolf 359 and Teegarden’s Star with values published in the literature. These stars were selected to span nearly the entire range of $L_{bol}$ values found in our sample.
\begin{table}[]
\centering
\caption{Comparison of bolometric luminosities ($L_{bol}$) derived with literature values for HD 197481, AD Leo, EV Lac, Proxima Centauri, Wolf 359 and Teegarden’s Star.}
\label{tab:tablbol}
\resizebox{\columnwidth}{!}{%
\begin{tabular}{ccc}
\hline\hline
\textbf{Star} & \textbf{$L_{bol}$ (Literature)} & \textbf{$L_{bol}$ (this work)} \\ \hline
HD 197481         & $1.02 \cdot 10^{-1} \pm 2 \cdot 10^{-3}$ $L_{\odot}$ $^{[1]}$              & $1.047 \cdot 10^{-1} \pm 6 \cdot 10^{-4}$ $L_{\odot}$              \\ 
AD Leo         & $2.35 \cdot 10^{-2} \pm 1.1 \cdot 10^{-5}$ $L_{\odot}$ $^{[2]}$              & $2.35 \cdot 10^{-2} \pm 8 \cdot 10^{-5}$ $L_{\odot}$              \\ 
EV Lac        & $1.28 \cdot 10^{-2} \pm 3 \cdot 10^{-4}$ $L_{\odot}$  $^{[3]}$            & $1.29 \cdot 10^{-2} \pm 6 \cdot 10^{-5}$ $L_{\odot}$             \\ 
Prox Cen      & $1.51 \cdot 10^{-3}\pm 8 \cdot 10^{-5}$ $L_{\odot}$   $^{[4]}$          & $1.51 \cdot 10^{-3} \pm 1.1 \cdot 10^{-5}$ $L_{\odot}$            \\ 
Wolf 359      & $1.06 \cdot 10^{-3}\pm 2 \cdot 10^{-5}$ $L_{\odot}$   $^{[5]}$          & $9.09 \cdot 10^{-4} \pm 5.1 \cdot 10^{-6}$ $L_{\odot}$            \\ 
Teegarden’s Star      & $7.22 \cdot 10^{-4}\pm 5 \cdot 10^{-6}$ $L_{\odot}$  $^{[6]}$           & $7.80 \cdot 10^{-4} \pm 5.7 \cdot 10^{-7}$ $L_{\odot}$            \\

\hline

\end{tabular}%

}
\tablefoot{[1] \cite{donati2023magnetic}, [2] \cite{bonfils2013harps}, [3] \cite{paudel2021simultaneous}, [4] \cite{ribas2017full}, [5] \cite{pineda2021m}, [6] \cite{dreizler2024teegarden}.}
\end{table}
\subsection{Flare properties}\label{flarefitting}
We based the flare identification technique on the method developed by \cite{colombo2022short}. The approach first identifies and removes long-term stellar variability from the light curve to highlight flares.  Subsequently, the procedure detects impulsive events exhibiting typical flare properties and calculates their amplitudes, timescales, and amounts of energy emitted.\\
We applied Gaussian Process (GP) fitting in an iterative way, and each time a residual curve (RC) was obtained. A 3$\sigma$ clipping procedure is applied to the residuals, continuing the process until the number of outliers is statistically consistent with expectations. We subtracted the GP prediction from the data at each iteration, using the Maximum A Posteriori (MAP) estimate obtained at step
$n$. By deliberately constraining the GP to avoid fitting short-timescale signals, this technique effectively smooths the data and ignores such transient features. The process is repeated until the number of data points beyond 3 $\sigma$ aligns with what is expected from the noise. The final zero-centered RC in which all small-scale events (i.e., the points removed by the iterative process) are eliminated is obtained by subtracting the last MAP forecast made by the GP from the original data.\\ The next step in our analysis consisted of identifying and fitting short-term activity events from the RC. We identified the maximum among all local maxima of the curve above 3 $\sigma$.  It is important to note that the final 3 $\sigma$ threshold differs from the earlier threshold applied to the residuals. The initial 3 $\sigma$ value is based on the statistical properties of the residuals after GP fitting, while the final 3 $\sigma$ is used to identify significant short-term events in the smoothed residual curve.
Next, we used a function composed of an increasing and a falling exponential to fit the events. The maximum of this function corresponds to the peak of event $A$ at $t_0$ (the peak's time coordinate):

\begin{equation}
    F(t)= H(t_0 - t)\cdot A e^{\frac{t-t_0}{t_r}} + H(t-t_0) \cdot A e^{\frac{t_0 -t}{t_d}}
\end{equation}

where $H(t_0 - t)$ is the Heaviside function, which equals zero for negative arguments and one for positive arguments, allowing the function to transition from increasing to decaying phase of the function. Here, $A$ is the flux at the peak of the event, while $t_r$ and $t_d$ are the rise and decay times, respectively, estimated by the fit.

\begin{figure}
    \centering
    \includegraphics[width=\hsize]{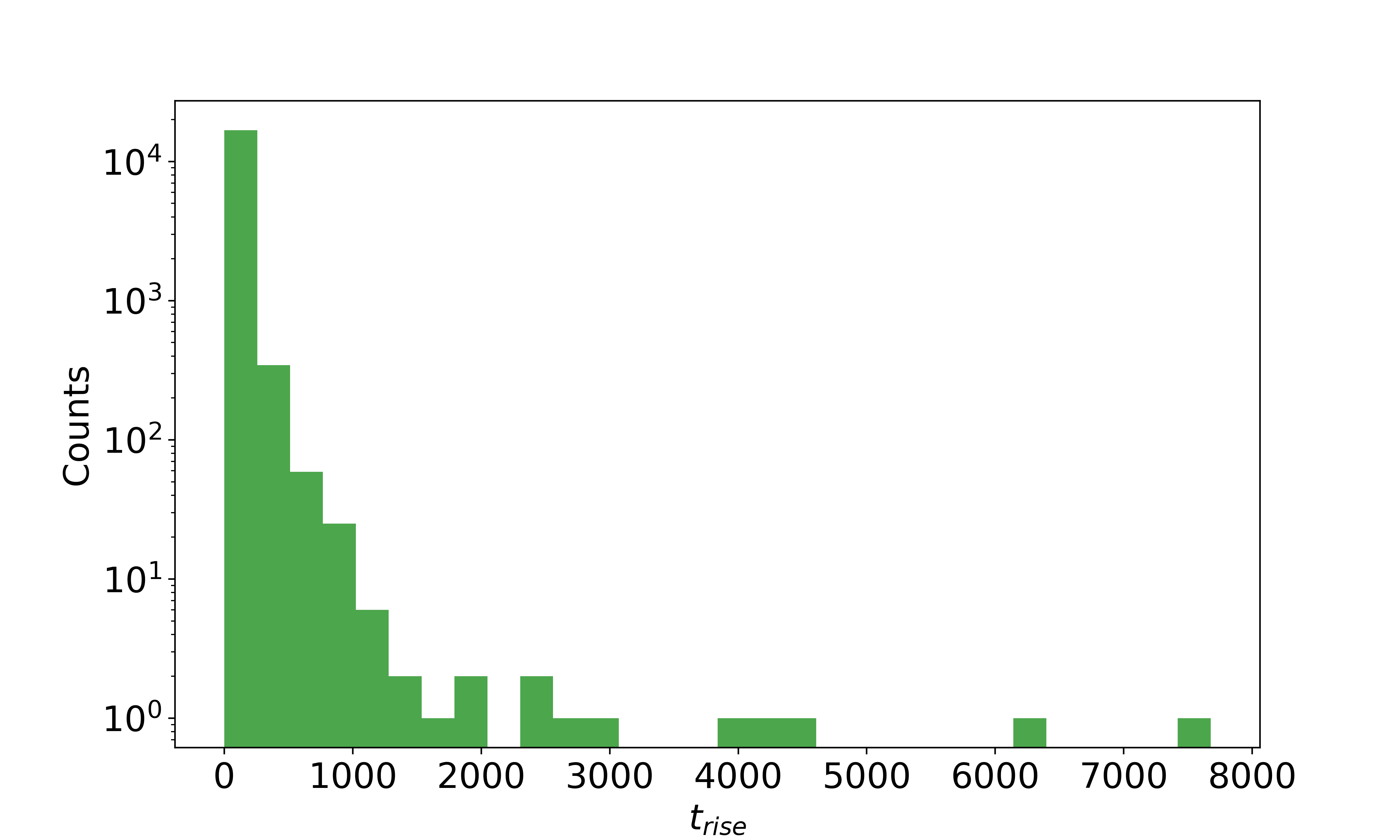}
    \caption{Distribution of $t_r$}
    \label{fig:trisehist}
\end{figure}

\begin{figure}
    \centering
    \includegraphics[width=\hsize]{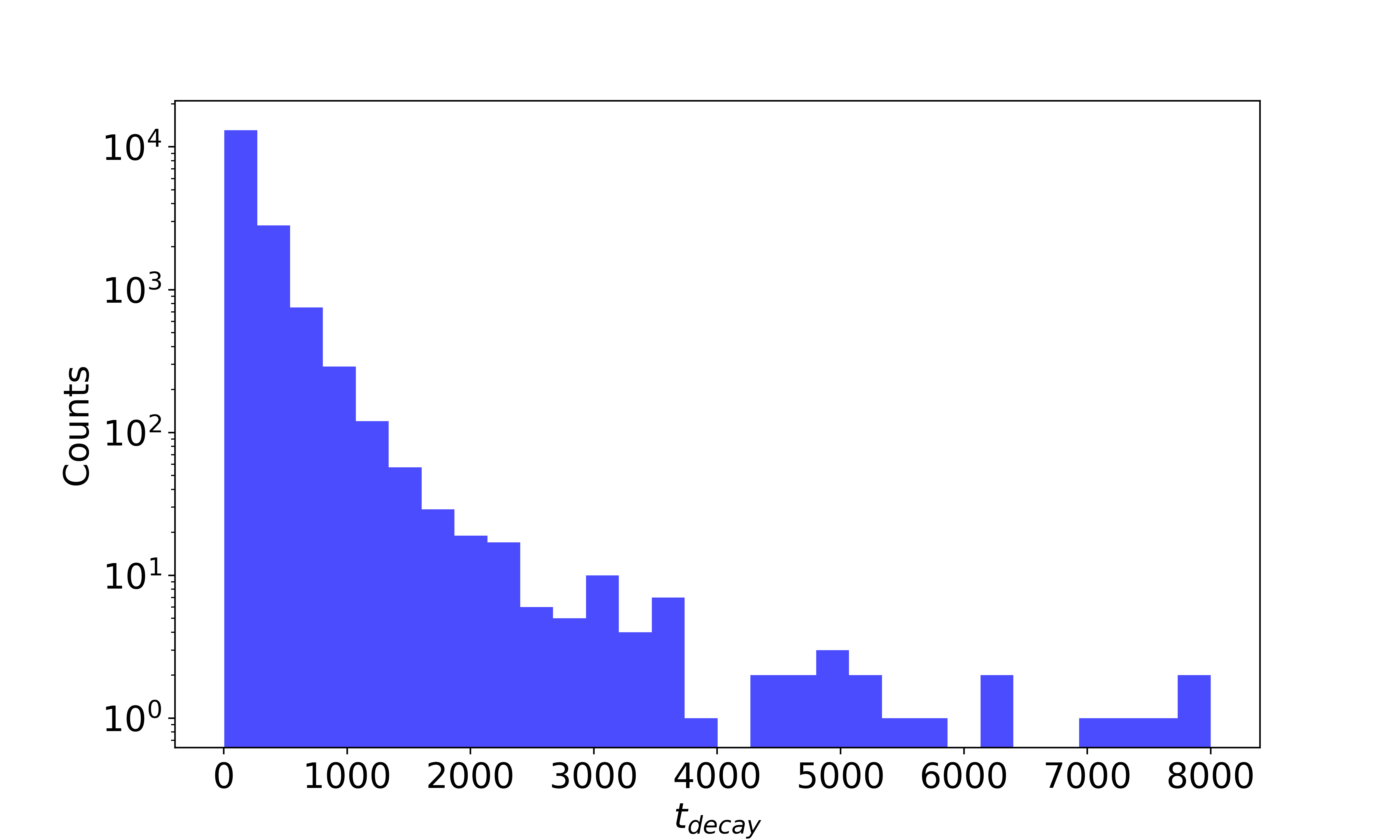}
    \caption{Distribution of $t_d$}
    \label{fig:tdecayhist}
\end{figure}

\begin{figure}
    \centering
    \includegraphics[width=\hsize]{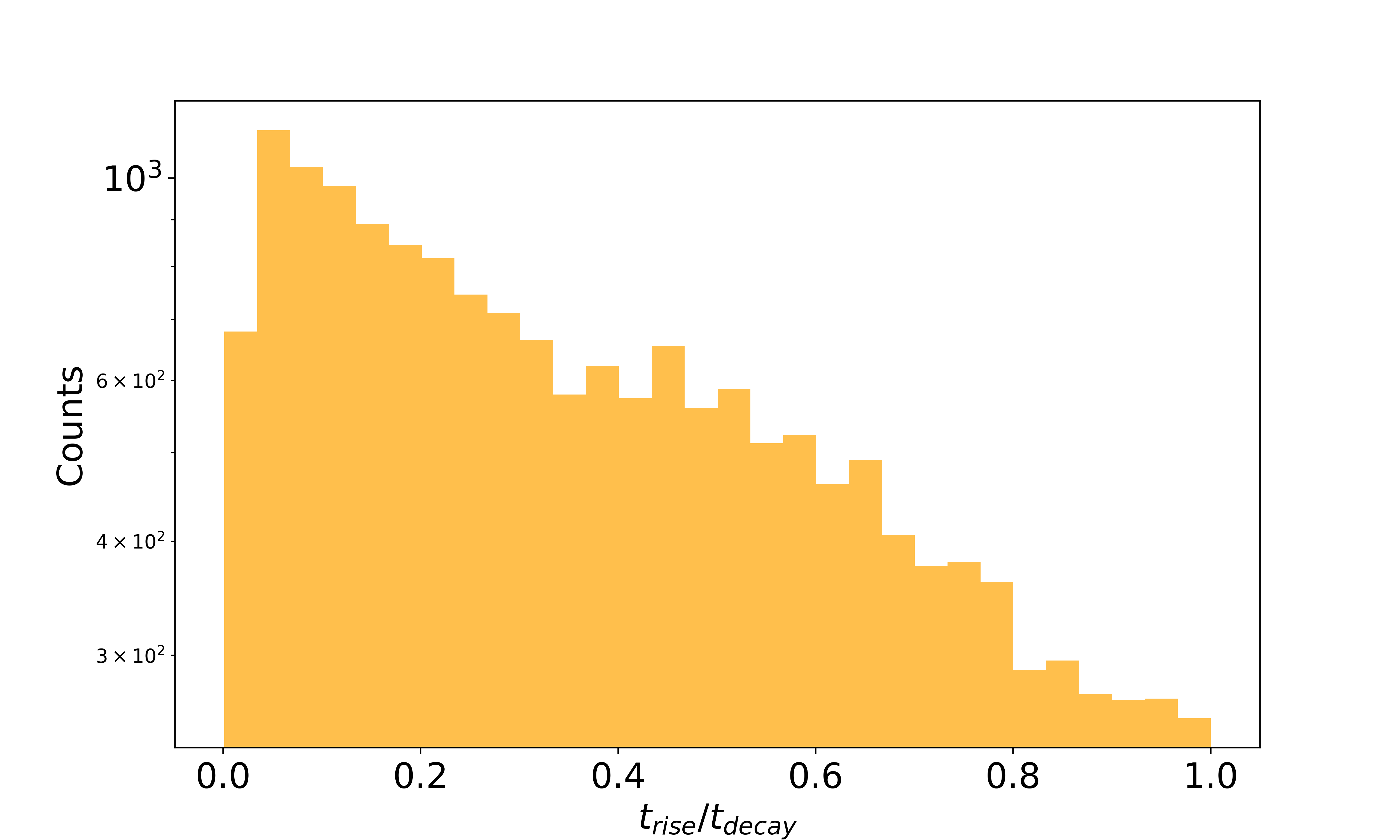}
    \caption{Distribution of $t_r/t_d$}
    \label{fig:tratiohist}
\end{figure}

Subsequently, we subtract the flare fits from the RC to highlight smaller energetic events overlapping the largest events, which will be identified through an iterative procedure. We fit all events until no more local maxima remain above the $3\sigma$ threshold. 
We modified the initial conditions and parameter constraints of the fitting procedure of \cite{colombo2022short} to make the code more general and applicable to a broader range of flares. In the original implementation, the number of free parameters was limited. Specifically, the peak time (P) and amplitude (A) were not included as free parameters, which could limit the flexibility of the fit.
To improve generalization, we introduced P and A as additional parameters, allowing the fit to better adapt to different flares. We constrained P within a narrow range around its initial estimate (P $\pm$ 10s) and imposed tight bounds on A (A $\pm$ 1\%) to prevent unphysical solutions. These modifications allow the fitting procedure to handle flares with a wider range of durations, amplitudes, and shapes while maintaining robustness in the parameter estimation.
We considered an event valid if the decay time exceeded the rise time, if the flare did not fall within a data gap larger than three times the TESS cadence (360 seconds) or smaller than the TESS cadence (120 seconds), and if the fitting procedure converged (we minimized $\chi^2$ with tolerance of $10^{-4}$).  Events that did not meet these criteria were discarded from subsequent analysis. \\
For each light curve, the analysis yielded the number of identified short-term events, their amplitudes and positions in time, their rise and decay times, and their energies. \\
We observed flares with minimum energies of $\sim 10^{29}$ erg and typical durations ranging from 2 to 8000 seconds. To investigate the timescales of these events, we analyzed the distributions of $t_{r}$, $t_{d}$, and their ratio ($t_{r} / t_{d}$). Fig. \ref{fig:trisehist} shows the distribution of $t_r$, revealing that most flares exhibit short rise times, with a few outliers extending to longer durations. In contrast, Fig. \ref{fig:tdecayhist} shows the distribution of $t_d$, has a broader and more evenly distributed range of values. Fig. \ref{fig:tratiohist} shows the distribution of $t_{r} / t_{d}$, which displays a decreasing trend, suggesting that most events have $\tau_r << \tau_d $.
We estimated the energy associated with each event by integrating the two fitted exponentials according to the following equation:

\begin{equation}
    E= (A \cdot t_r + A \cdot t_d)\frac{L_{bol}}{L_{TESS}},
\end{equation}
where $L_{bol}$ is the bolometric luminosity of the star and $L_{TESS}$ is the luminosity in the TESS bandpass.\\
TESS observes stars within a specific wavelength range of 600-1000 nm, meaning that our energy estimates for stellar flares may not capture the full emission, particularly in the UV and X-ray bands. Although this leads to an underestimation of the total flare energy, the TESS band still provides a reliable measure for our analysis. For our study, these estimates were sufficient to study flare properties, and further multiwavelength observations would serve only to refine the values.
Hereafter,  we use the term amplitude to denote A (the peak of the event) and $\tau$ to refer to $t_r+t_d$ indicating the duration of the flare.

\subsection{Cumulative frequency energy distribution}\label{cumulative}
For every star in our dataset, we created a cumulative flare energy distribution, representing the total number of flares above each energy value.\\
The cumulative energy flare distribution in the log-log representation can be parameterized by a power law. In practice, at low energy, the number of detected flares can be largely incomplete, causing a flattening of the curve that therefore needs to be parameterized by two slopes. The first slope is defined in the high-energy regime, where the flare detection can be considered complete, and the second in the low-energy range, where the results are incomplete. The energy break corresponding to the transition between the two regimes depends on the sensitivity of the given observations. This transition is not at a precise energy, since our detection is based on the flare peak amplitude, while energy depends on both amplitude and flare decay time. Consequently, we can detect flares with large amplitude and short decay time, but flares with low amplitude and long decay time may remain undetected.

\subsection{Slope and energy break calculations}\label{slopecalculations}
We developed a tool to determine the slopes and the energy break of the cumulative curve derived in the previous subsection. \\
The tool uses an approach inspired by the analysis of fast radio bursts (FRBs), as reported in studies by \cite{aggarwal2021comprehensive} and \cite{hewitt2022arecibo}, to calculate the energy break in stellar flare data. We computed weights for the fitting procedure on the basis of the uncertainty in the logged flare numbers.\\
The log-transformed flare energy and flare number data are fitted piecewise linearly using the PiecewiseLinFit Python package. This method identifies breakpoints where the slope of the relationship between log energy and log number changes. The script captures variations in the slope at various energy levels using a two-segment fit (one breakpoint).\\ We define the term "energy break" as the energy value corresponding to the point where the curves change the slope. If fewer than six flares are present, we use a single-segment fit and do not include breakpoints.\\
The script uses a bootstrap sampling technique to ensure the robustness of the identified energy break. It creates random samples inside the data points' confidence intervals and fits each sample using piecewise linear fitting. This yields a distribution of breakpoints from which the slopes, intercepts, and mean and standard deviation of the breakpoints are computed. These statistical metrics provide an estimate of the energy break's uncertainty.
For each sector, the script performs additional checks to optimize the fit. If there are fewer than six flare events, it saves the single slope, as no energy break exists in a single-segment linear fit, and does not record an energy break. If the number of flares after the energy break is less than or equal to two, it saves the first slope from the double fit, treating the first segment as valid, while ignoring the second slope and not recording an energy break. If neither condition applies, it saves the second slope from the double, treating the second segment as valid and also records the energy break. Additionally, the fitted values for energies greater than the energy break are extracted and saved for further analysis.

\begin{figure*}
    \centering
    \subfloat[]{
        \includegraphics[width=0.4\textwidth]{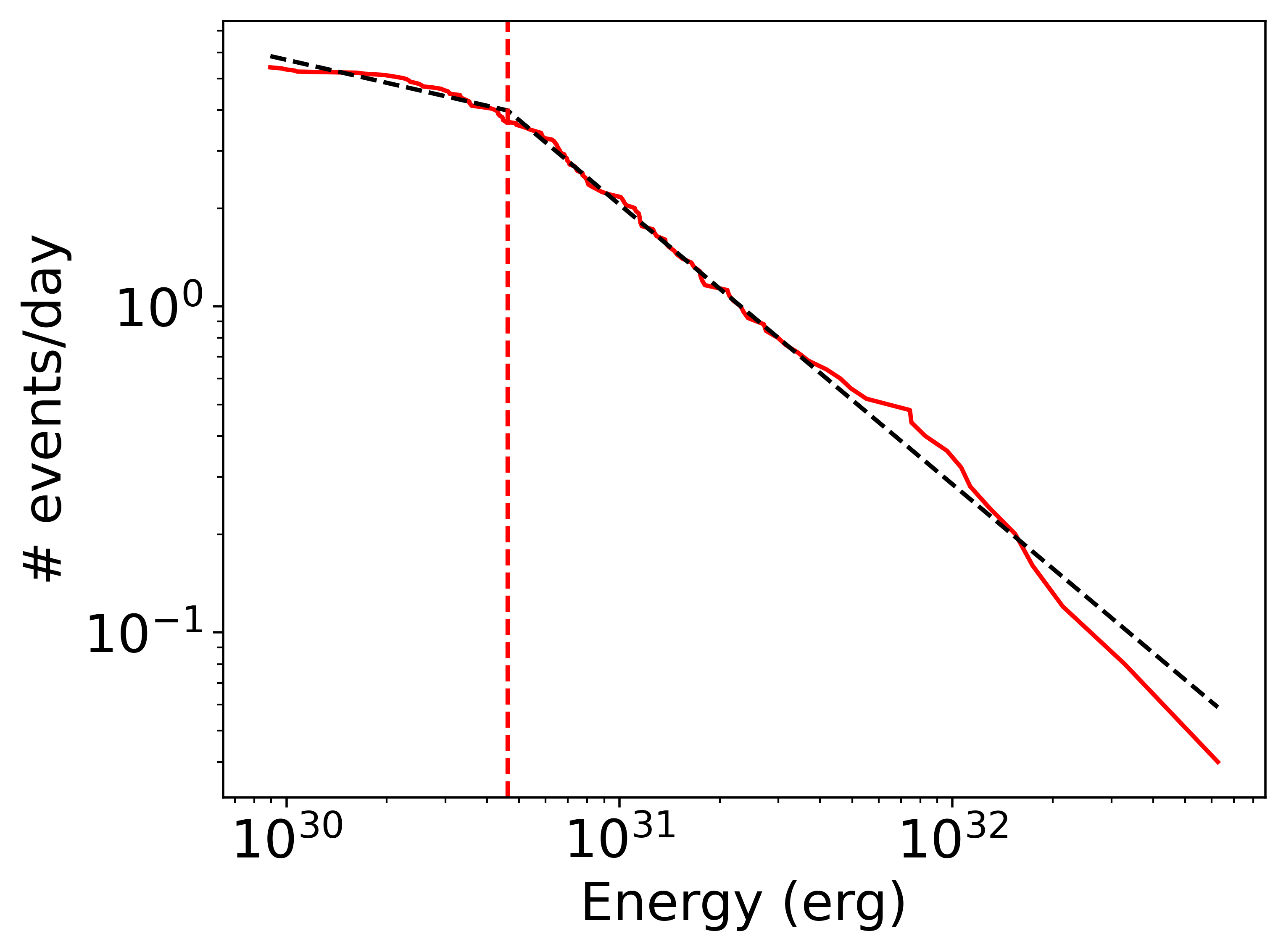}
        \label{fig:twosegment}
    }
    \subfloat[]{
        \includegraphics[width=0.4\textwidth]{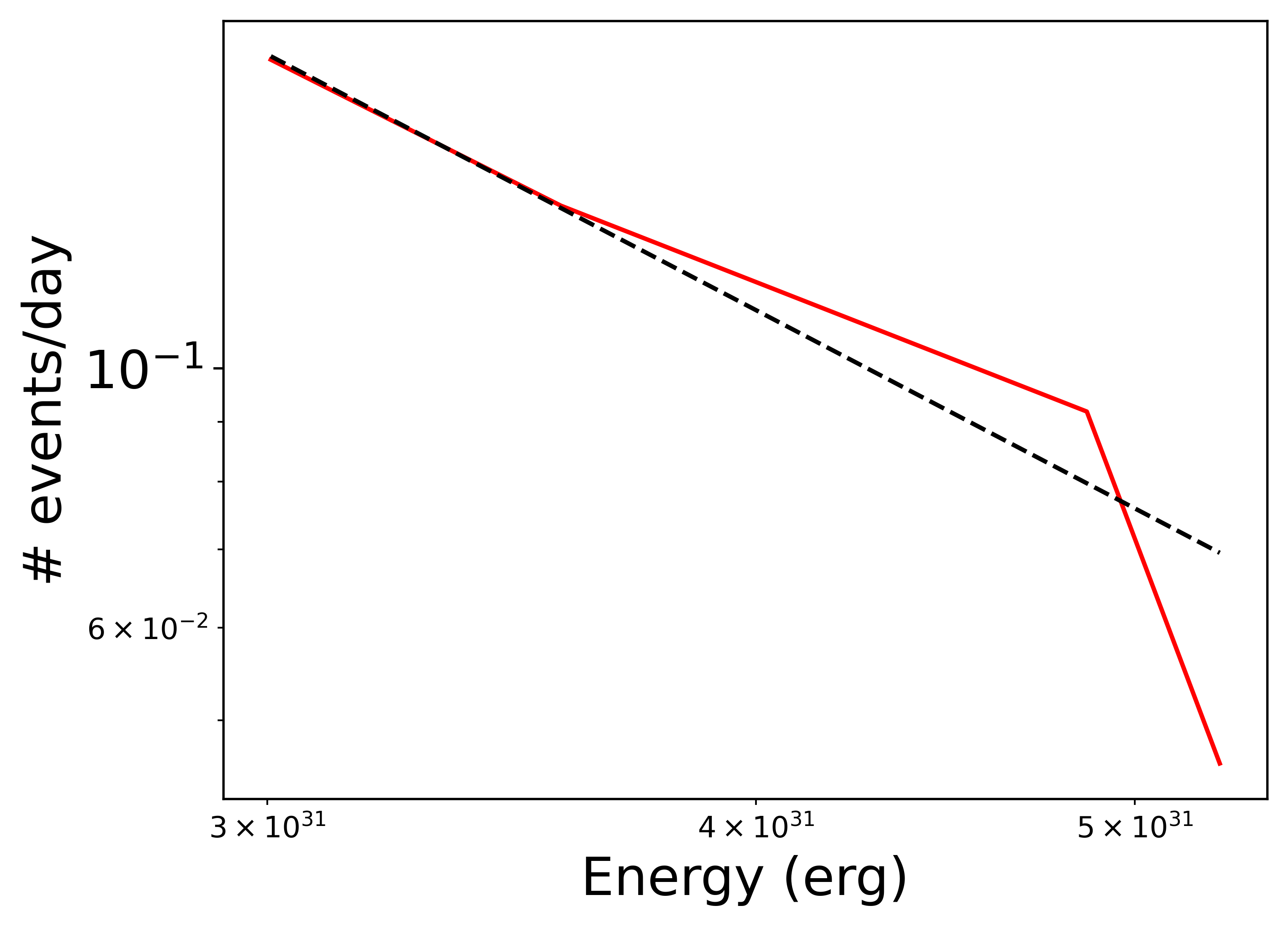}
        \label{fig:onesegment}
    }
    \caption{Examples of the fitting method: (a) Two-segment fit. The red line is the example of a cumulative curve of a star in a given sector, the black dashed lines represent the fit and the red dashed vertical line represents the break point. (b) Analogous to panel (a), showing a one-segment fit.}
    \label{fig:segments}
\end{figure*}

Globally, 651 sectors fall under case number 3, while 100 sectors correspond to cases 1 or 2.
The additional checks described above ensure that the fitting process is robust and that the results are reliable, even with limited data points.\\ 
Figure \ref{fig:twosegment} shows an example of the two-slope fit. The red line is the cumulative curve, the black dashed lines indicate the fit, and the red dashed vertical line marks the breakpoint. Figure \ref{fig:onesegment} illustrates an example of the one-slope fit.\\
When the slope exceeds the critical value of -1, the distribution is dominated by high-energy flares, indicating that the majority of the flare energy arises from the most energetic events. For lower values, the distribution tends to diverge at low energy, implying the need for the curve to flatten below a certain energy threshold.
\section{Results}\label{results}
The main results from our study on stellar activity and flare properties in our sample of stars are presented in this section. We detected 17,229 flares, with the number of flares per sector ranging from a minimum of zero to a maximum of 76. 
These figures highlight the considerable variability in stellar activity within our sample, demonstrating that some stars exhibited frequent flare events, while others showed little to no activity. This range in flare occurrences is crucial for understanding differences in stellar behavior and provides context for comparing activity levels across the different stars in our study. In subsection \ref{flarefrequency} we focus on the frequency and energy of stellar flares observed in our sample. To illustrate the range of flare activity from least active to most active stars, we present three representative cases. Subsection \ref{slope} delves into the analysis of the slopes and energy breaks in our sample, aiming to identify any patterns or connections between the slope and stellar properties.\\ Finally, in \ref{manysectors},  we conduct a detailed investigation of G 227-22 and G 258-33, the stars with the largest number of observational sectors. The extensive data coverage for these stars allows for an in-depth analysis of their flare activity.

\subsection{Flare frequency and energy}\label{flarefrequency}
We performed a detailed analysis of the energy distribution of the events. Table 3 reports, for each star and sector, the number of detected flares, the $L_{bol}$, the slope, and the break energy for cases fitted with a two-slope model. We present detailed results for three representative stars in our analyzed sample.  BD-156290 is a quiet star with a small number of flares, averaging $\leq$ 10 flares per sector. YZ CMi is a moderately active star with a medium/high number of flares, with an average flare per sector of $\geq$ 20. EV Lac is a highly active star with a large number of flares, averaging $\geq$ 40 flares per sector.

\begin{figure*}
    \centering
    \subfloat[]{
        \includegraphics[width=0.32\hsize]{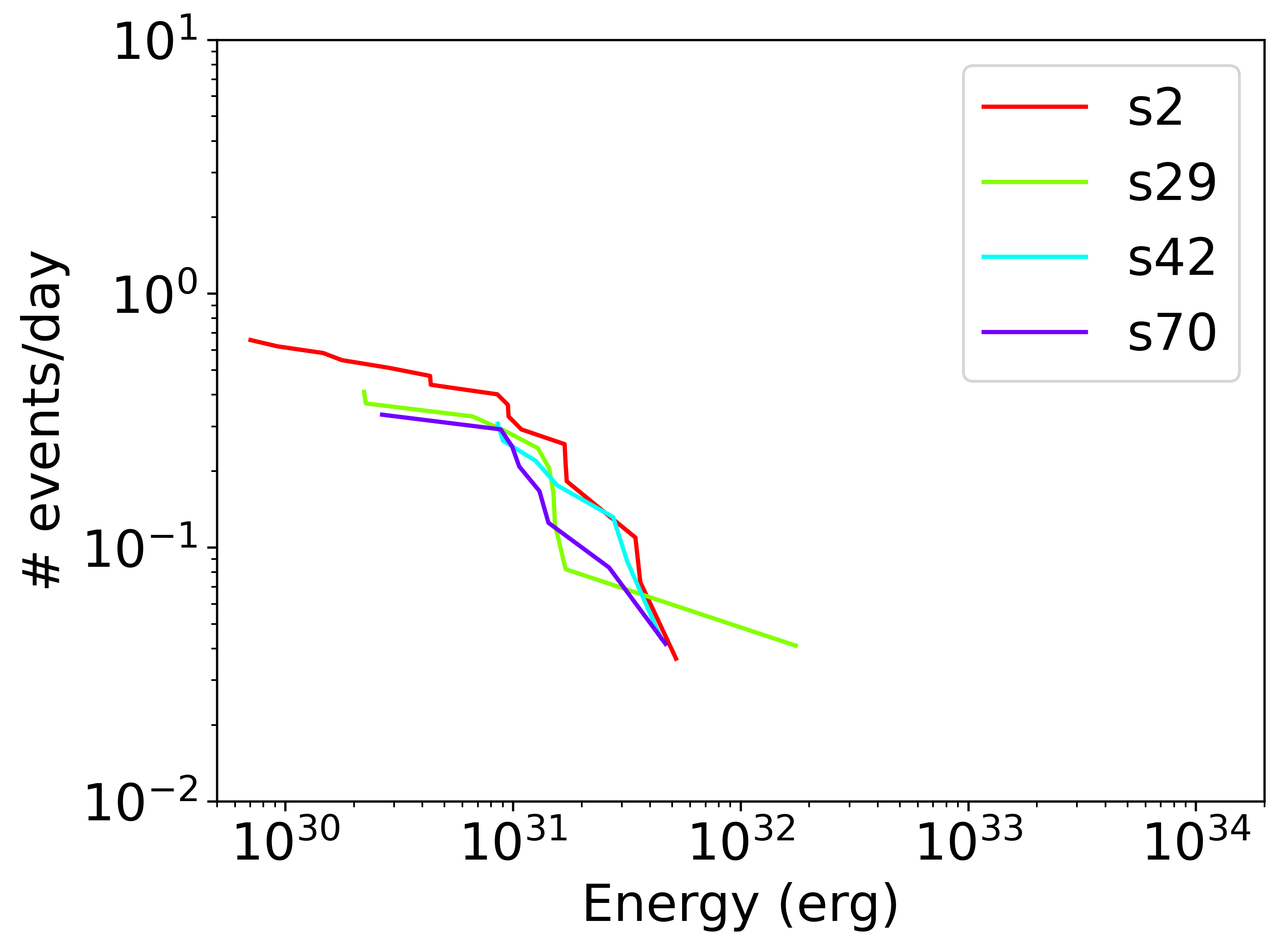}
        \label{fig:flarebd}
    }
    \subfloat[]{
        \includegraphics[width=0.32\hsize]{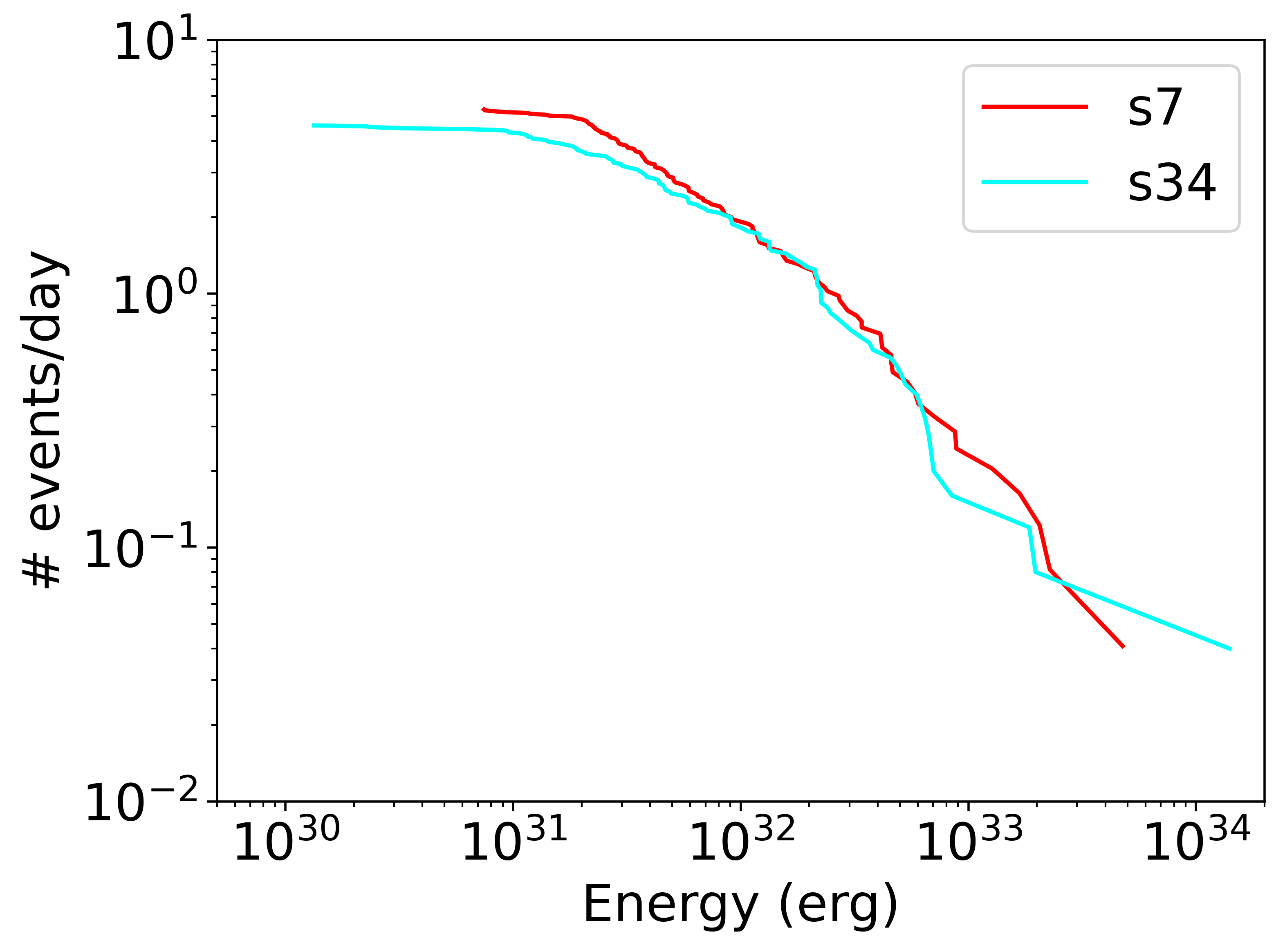}
        \label{fig:flarecmi}
    }
    \subfloat[]{
        \includegraphics[width=0.32\hsize]{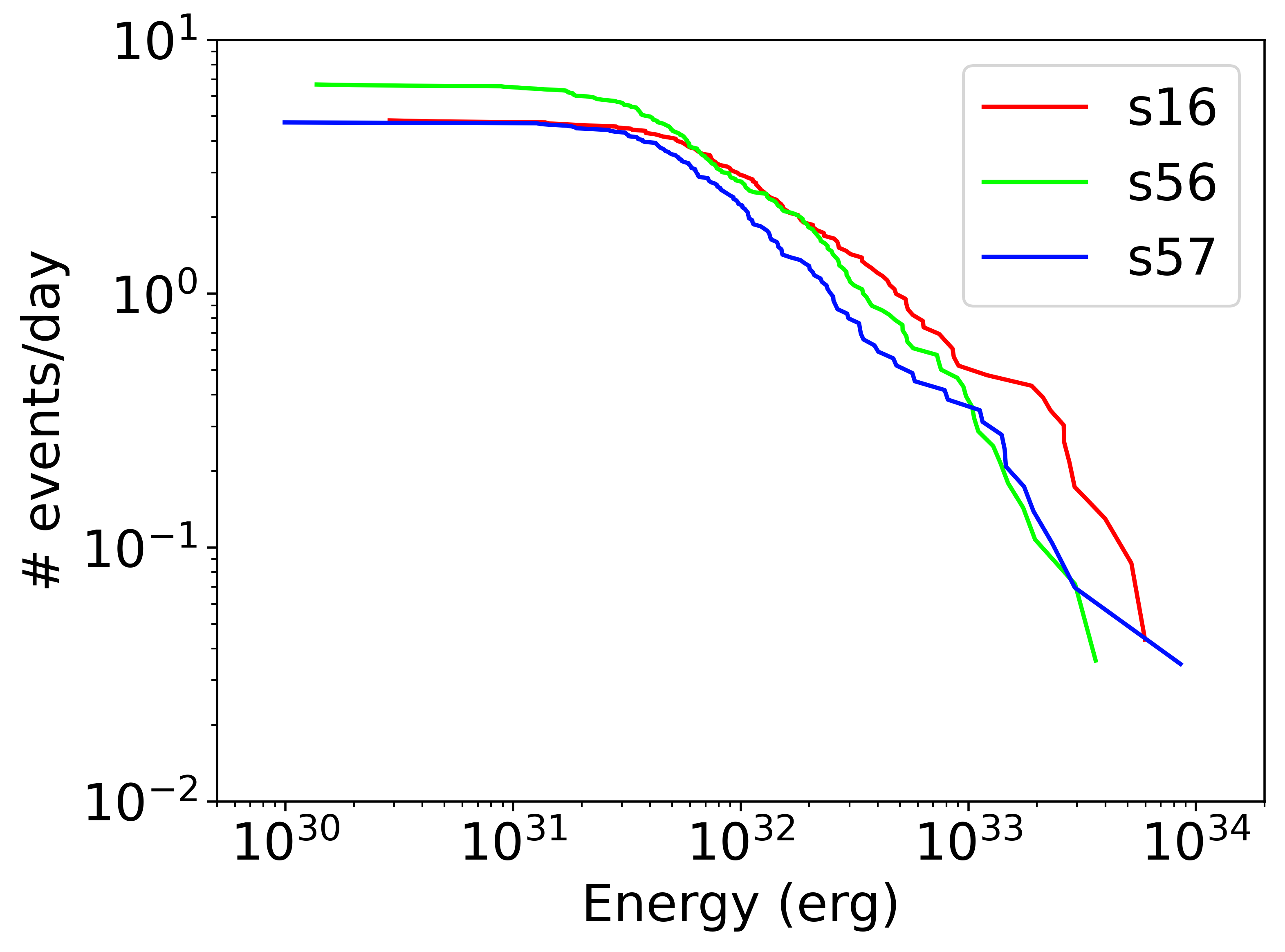}
        \label{fig:flareevlac}
    }
    \caption{Cumulative curve showing events per day vs energy in each sector. From left to right: BD-156290 (panel a), YZ CMi (panel b), and EV Lac (panel c).}
    \label{fig:flare_all}
\end{figure*}

\begin{figure*}
    \centering
    \subfloat[]{
        \includegraphics[width=0.32\hsize]{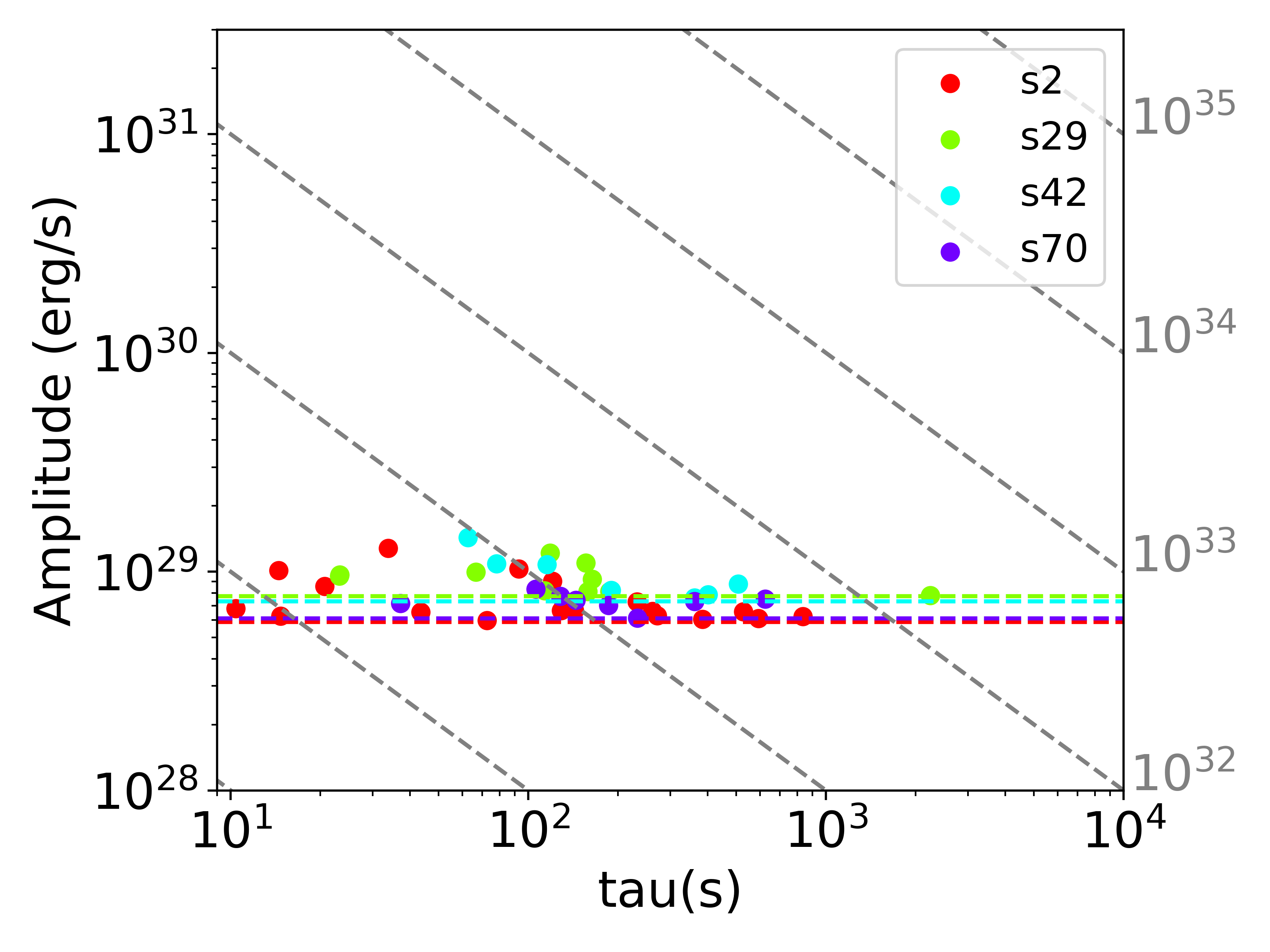}
        \label{fig:amplitudebd}
    }
    \subfloat[]{
        \includegraphics[width=0.32\hsize]{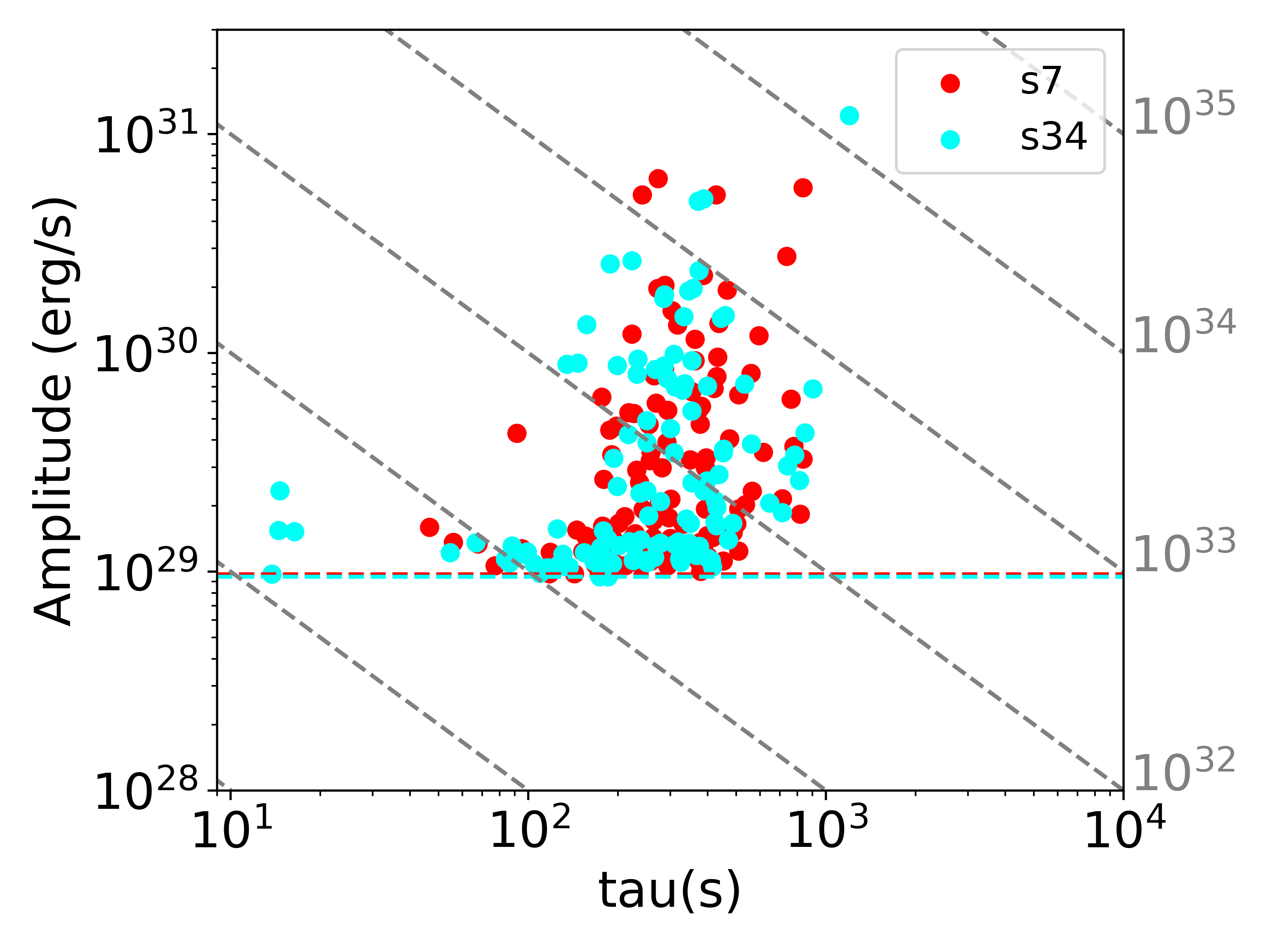}
        \label{fig:amplitudecmi}
    }
    \subfloat[]{
        \includegraphics[width=0.32\hsize]{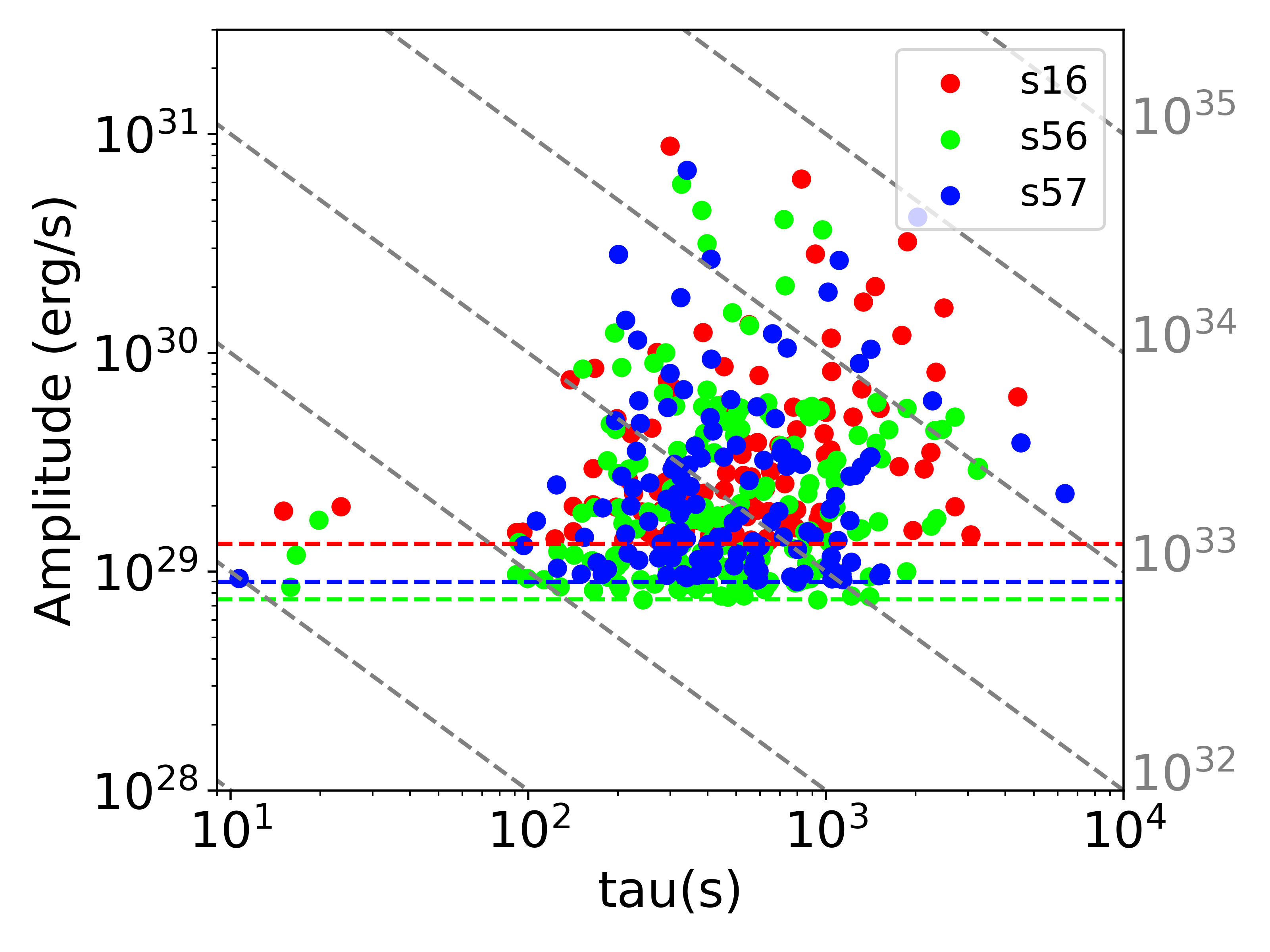}
        \label{fig:amplitudeevlac}
    }
    \caption{Amplitude vs $\tau$ for events in each sector. The colored horizontal lines indicate the value of 3$\sigma$ used to discriminate events from noise, while the gray oblique lines represent the isoenergetic lines. From left to right: BD-156290 (panel a), YZ CMi (panel b),  and EV Lac (panel c).}
    \label{fig:amplitude_all}
\end{figure*}

These three stars are directly comparable because they have very similar $L_{bol}$. For BD-156290, $L_{bol}=0.01291\, L_{\odot}$; for YZ CMi, $L_{bol}=0.01137\, L_{\odot}$; and for EV Lac, $L_{bol}=0.0128\, L_{\odot}$.
Figures \ref{fig:flarebd}, \ref{fig:flarecmi}, and \ref{fig:flareevlac} display the cumulative flare frequency as a function of energy for each star. In Fig. \ref{fig:flarebd}, BD-156290 shows approximately 0.4 flares per day with energies $\geq 1 \times 10^{31}$ erg. Figure \ref{fig:flarecmi} illustrates that YZ CMi experienced about five daily flares at the same energy threshold. For EV Lac, shown in Fig. \ref{fig:flareevlac}, the star exhibited between 6 and 7 daily flares, but at a higher energy level of $\geq 1 \times 10^{32}$ erg.\\
For BD-156290, the cumulative flare frequency distribution yielded a mean slope of $-1.32 \pm 0.55$ and an energy break of $1.16 \cdot 10^{31}$ erg. Similarly, YZ CMi exhibited a slope of $-0.92 \pm 0.27$ with an energy break of $8.56 \cdot 10^{32}$ erg, while EV Lac showed a slope of $-0.79 \pm 0.06$ with an energy break of $5.60 \cdot 10^{31}$ erg.\\
We observed that stars with high flare activity, such as EV Lac or YZ CMi, display flatter energy distributions and higher energy thresholds, indicating their capacity to produce frequent high-energy flares. In contrast, less active stars, such as BD-156290, with steeper slopes and lower break energies, predominantly release energy in less powerful flares. This reflects a difference in the structure and strength of their magnetic fields, with more active stars likely having stronger and more complex magnetic fields, capable of storing and impulsively releasing greater energy.\\
As part of this analysis, and to understand the origin of the slope changes with activity level, we present plots of the flare amplitude as a function of the $\tau$ for the identified events in Fig. \ref{fig:amplitudebd}, \ref{fig:amplitudecmi}, \ref{fig:amplitudeevlac}.
In these figures, dotted colored lines represent the thresholds used for event selection, while gray oblique lines indicate isoenergetic lines, providing information on sensitivity and completeness of the analysis. 
The sensitivity of the observations for the three stars was very similar, making the results directly comparable.
Our procedure identified events with decay times ranging from 100 seconds to several thousand seconds. 
Comparing the three cases, we found that the quietest star, BD-156290, exhibited short flares of small amplitude. The star with intermediate activity, YZ CMi, showed equally short flares but with amplitudes nearly two orders of magnitude higher. In contrast, EV Lac— one of the most active stars in our sample—displayed flares with amplitudes comparable to those of YZ CMi but with longer decay times, up to 5000 seconds, making them more energetic.

\subsection{Slope and energy breaks analysis}\label{slope}
In this subsection, we focus on analyzing the distributions of the energy breaks and slopes for the entire sample.
Among the 751 sectors examined, 651 were fitted with two-segment models, while 100 were fitted with single-segment models. \\
For sectors fitted with two-segment models, the mean slope was -1.23,  with a standard deviation of 1.32 and a median slope of -0.93. In contrast, the single-segment model fits yielded a mean slope is -0.72, a standard deviation is 0.64 and a median slope of -0.53.
Table \ref{tab:tabslopesmean} summarizes the mean slope values and the number of sectors for each fitting method. The slopes derived from one- and two- slopes are fully consistent.

\begin{table}[]
\centering
\caption{Mean slopes and number of sectors for the two fitting methods.}
\label{tab:tabslopesmean}
\resizebox{\columnwidth}{!}{%
\begin{tabular}{ccc}
\hline\hline
\textbf{Method} & \textbf{Slopes mean $\pm$ std} & \textbf{\# sectors}  \\ \hline
Two-segment fit & -1.23 $\pm$ 1.32               & 651                  \\ 
One-segment fit & -0.79 $\pm$ 0.64               & 100                          \\ 
\hline
\end{tabular}%
}
\end{table}

\subsection{Flaring stars with long observations}\label{manysectors}
\begin{figure*}
    \centering
    \subfloat[]{
        \includegraphics[width=0.45\hsize]{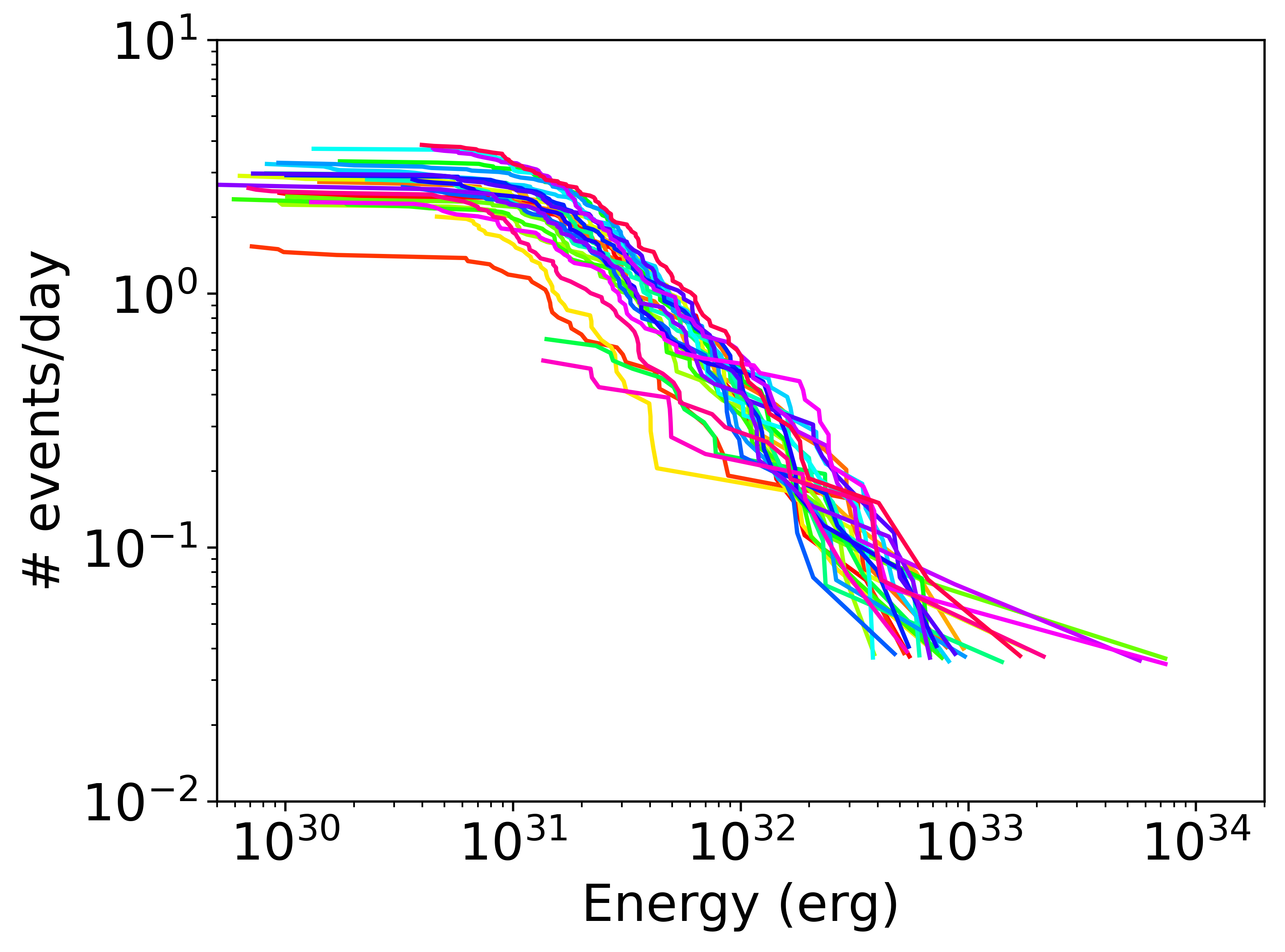}
        \label{fig:flareG227}
    }
    \subfloat[]{
       \includegraphics[width=0.45\hsize]{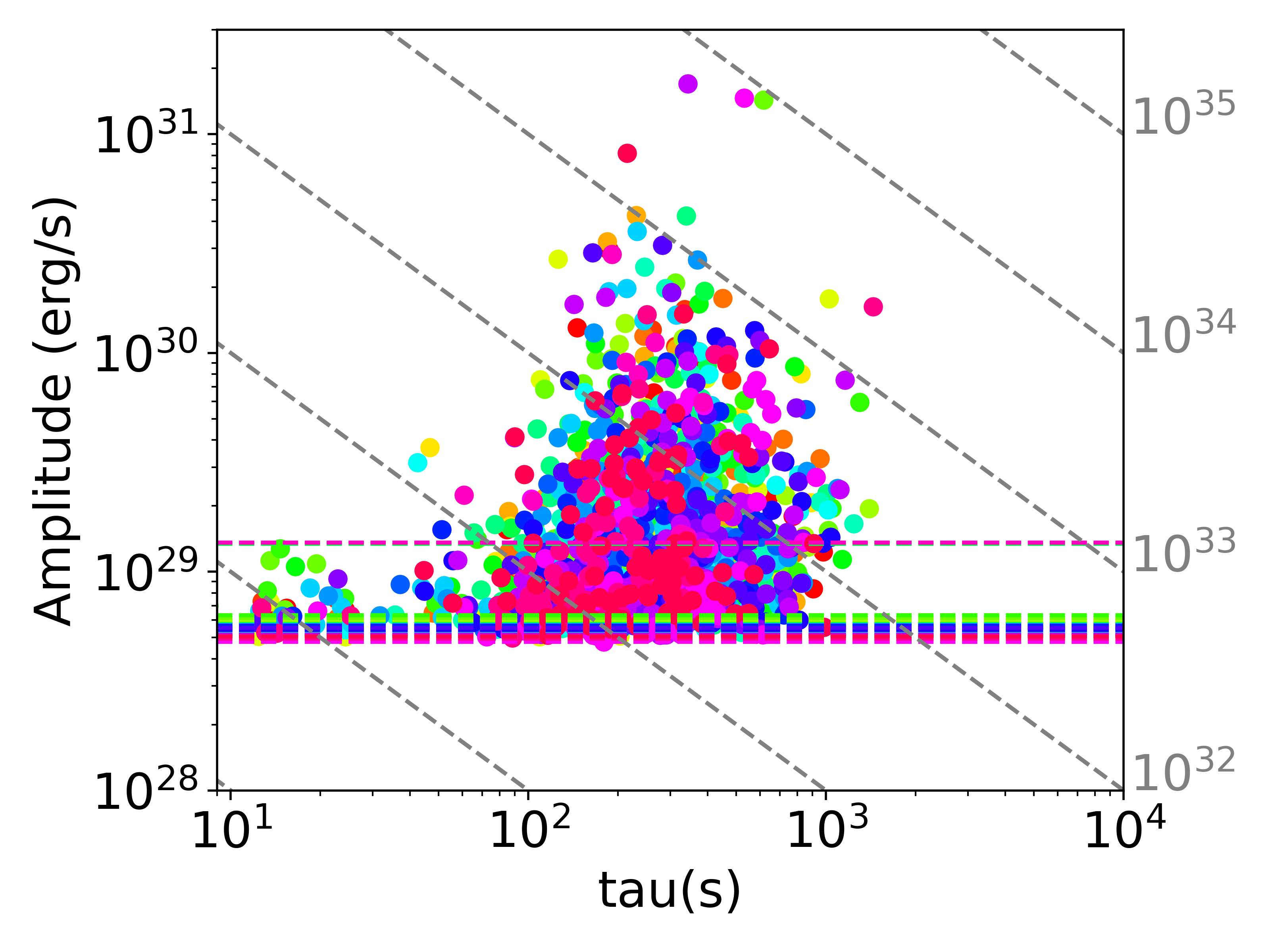}
        \label{fig:amplitudeG227}
    }
    \caption{(a) Cumulative events per day vs energy for each sector of G 227-22. Between 2 and 4 events per day are detected with energies $\geq 10^{31}$ erg. (b) Amplitude vs $\tau$ for events in G 227-22. The colored horizontal lines indicate the value of 3$\sigma$ used to discriminate events from noise, while the gray oblique lines represent isoenergetic lines.}
    \label{fig:bothG227}
\end{figure*}

\begin{figure*}
    \centering
    \subfloat[]{
        \includegraphics[width=0.45\hsize]{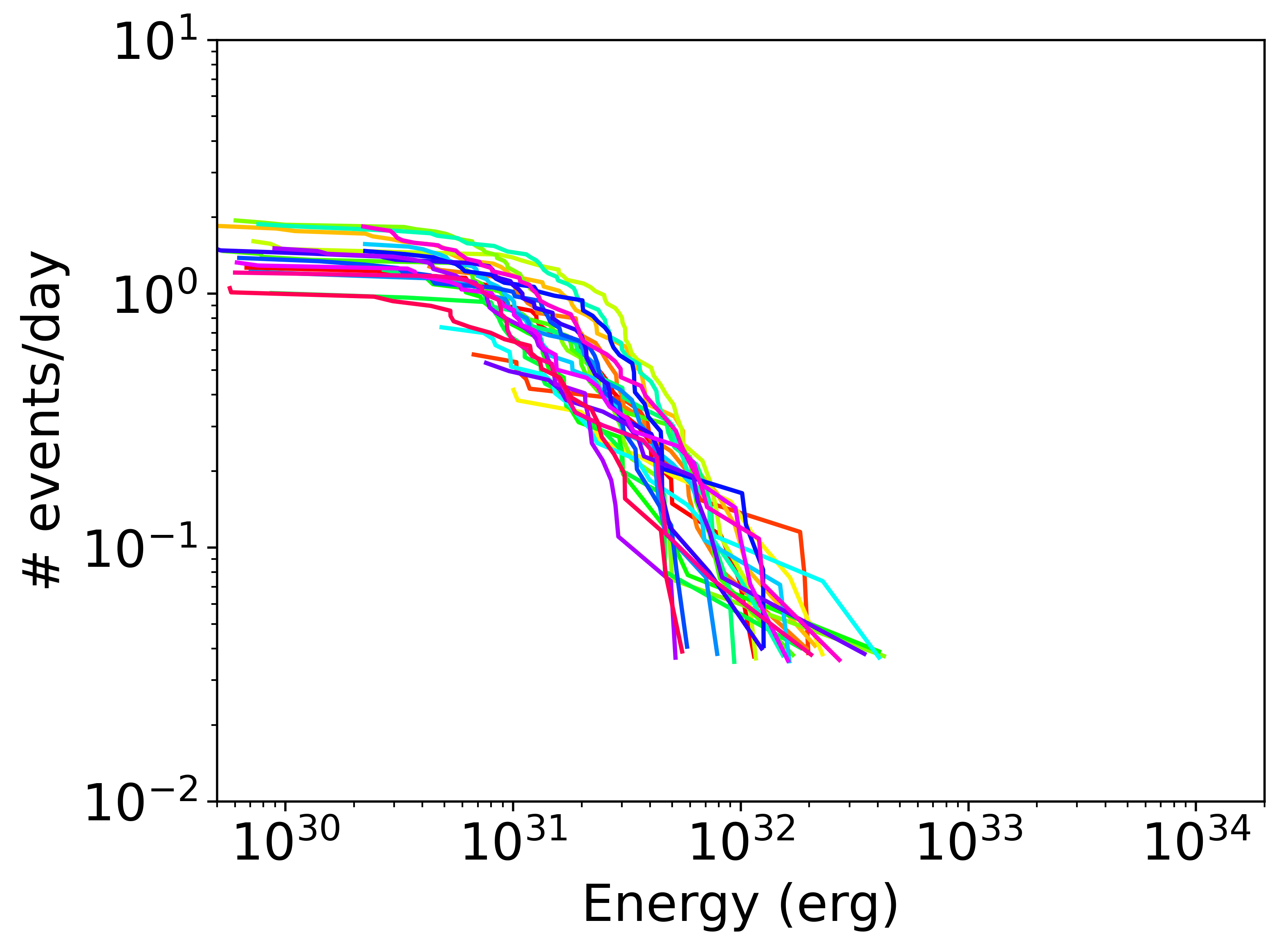}
        \label{fig:flareG258}
    }
    \subfloat[]{
        \includegraphics[width=0.45\hsize]{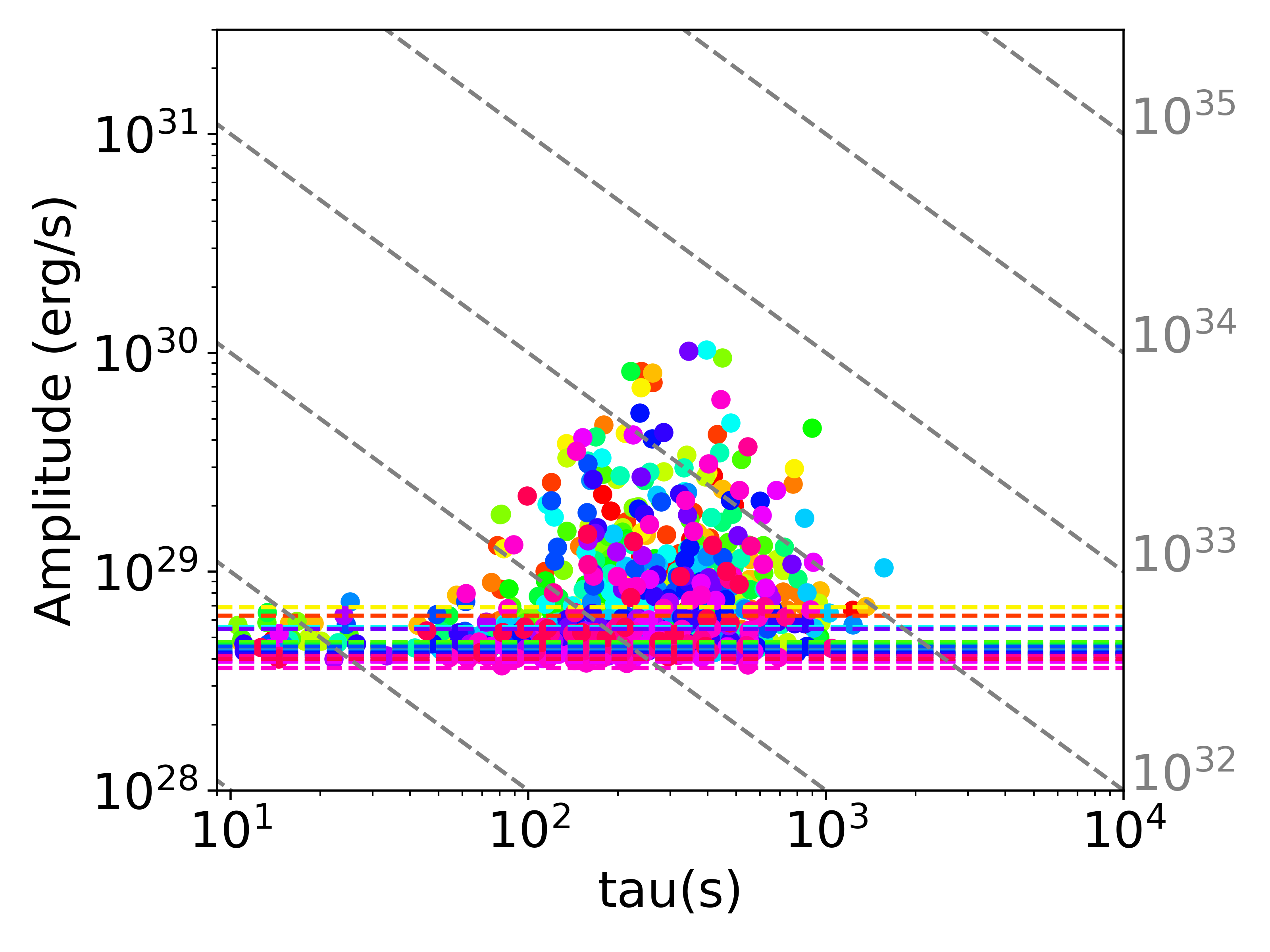}
        \label{fig:amplitudeG258}
    }
    \caption{(a) Cumulative events per day vs energy for each sector of G 258-33. Between 0.6 and 2 events per day are detected with energies $\geq 10^{31}$ erg. (b) Amplitude vs $\tau$ for events in G 258-33. The colored horizontal lines indicate the value of 3$\sigma$ used to discriminate events from noise, while the gray oblique lines represent the isoenergetic lines.}
    \label{fig:bothG258}
\end{figure*}

\begin{figure*}
    \centering
    \subfloat[]{
        \includegraphics[width=0.45\hsize]{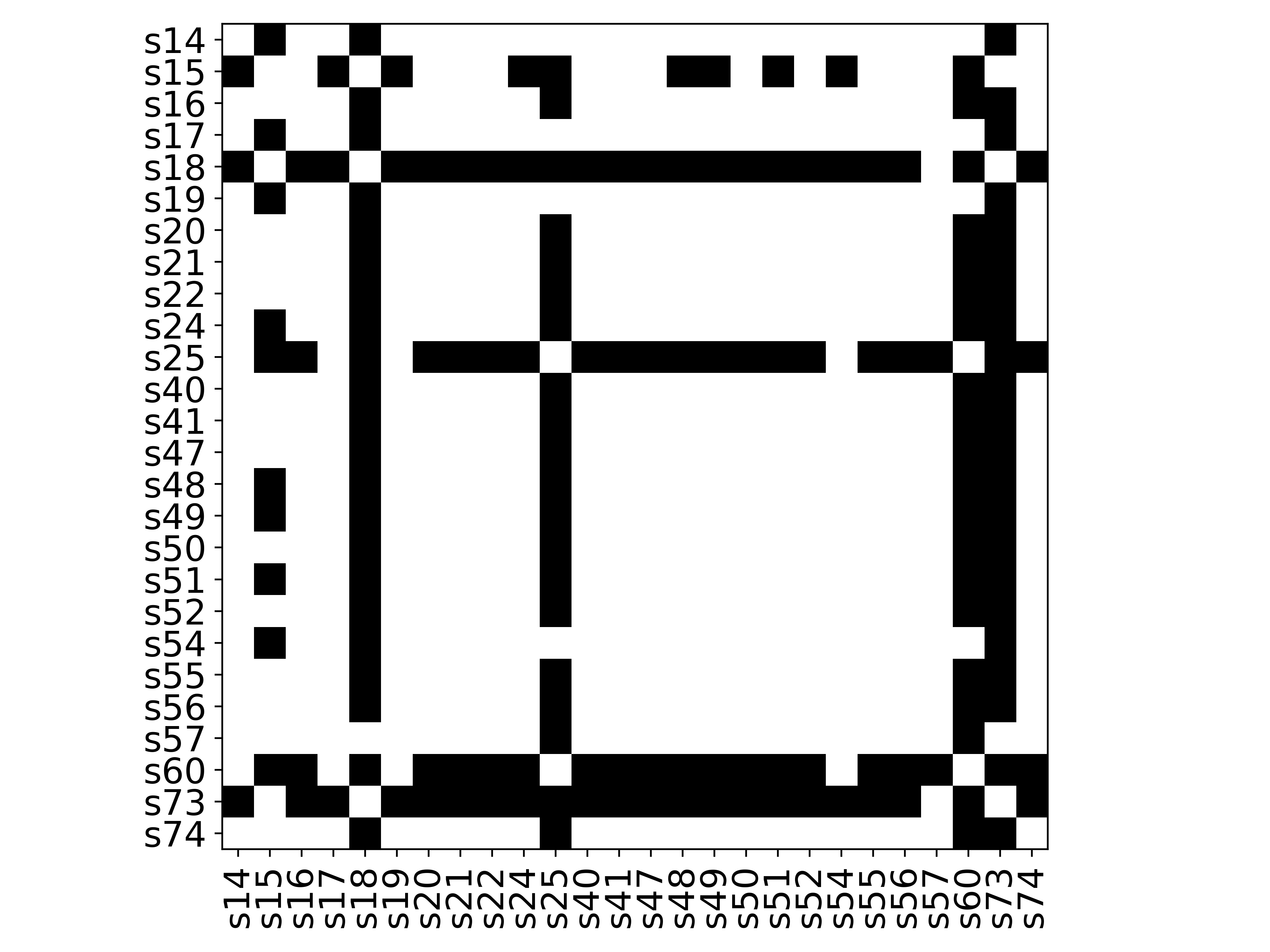}
        \label{fig:matrixG227}
    }
    \subfloat[]{
       \includegraphics[width=0.45\hsize]{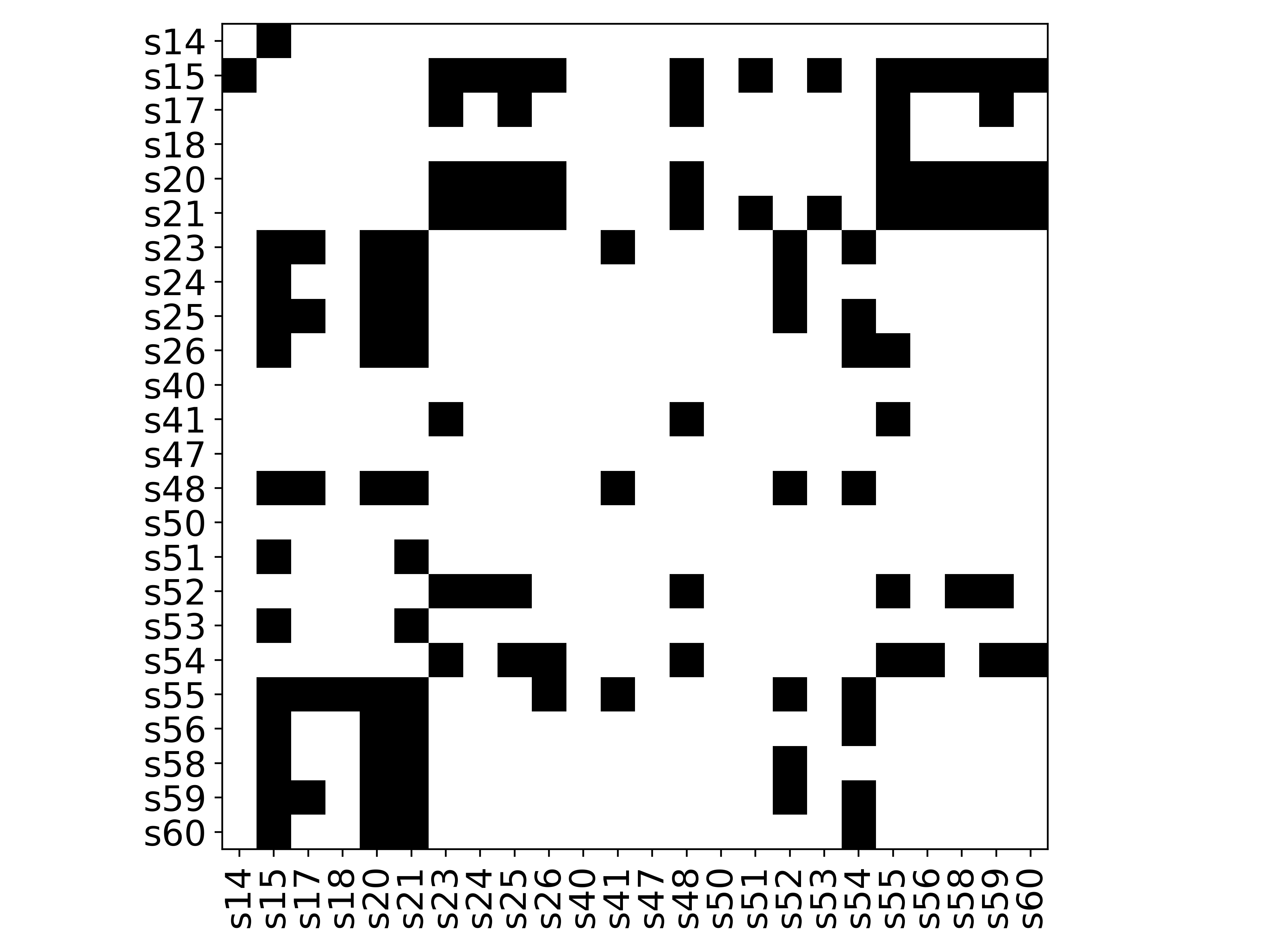}
        \label{fig:matrixG258}
    }
    \caption{Comparison matrices between sector pairs of stars G 227-22 and G 258-33. (a) Matrix for G 227-22; black cells indicate sector pairs that are not comparable (b) Same as panel (a), but for G 258-33.}
    \label{fig:bothMatrices}
\end{figure*}

\begin{figure}
    \centering
    \includegraphics[width=\hsize]{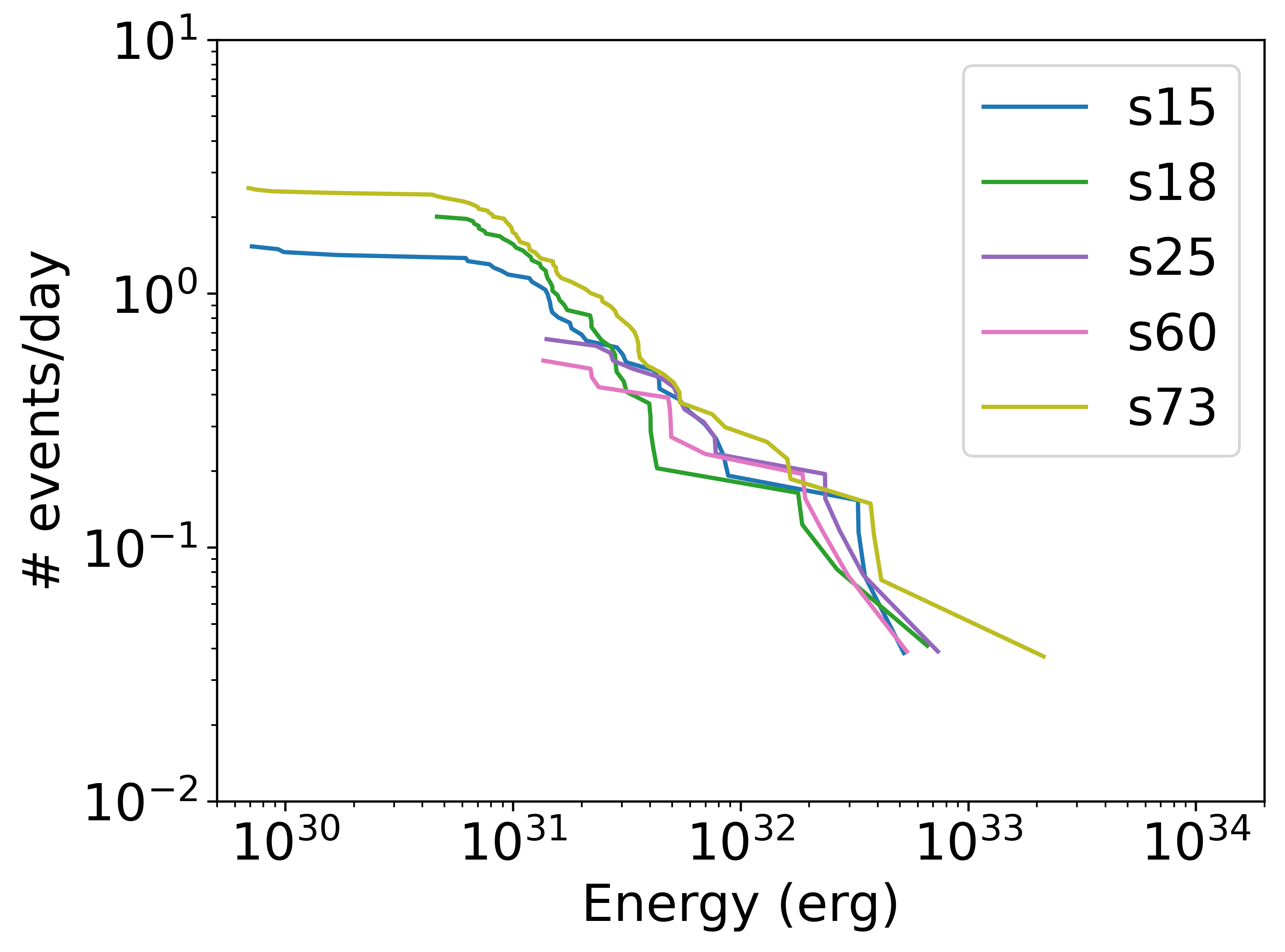}
    \caption{Cumulative events per day vs energy for non-comparable sectors of G 227-22.}
    \label{fig:flareG227ks}
\end{figure}

\begin{figure}
    \centering
    \includegraphics[width=\hsize]{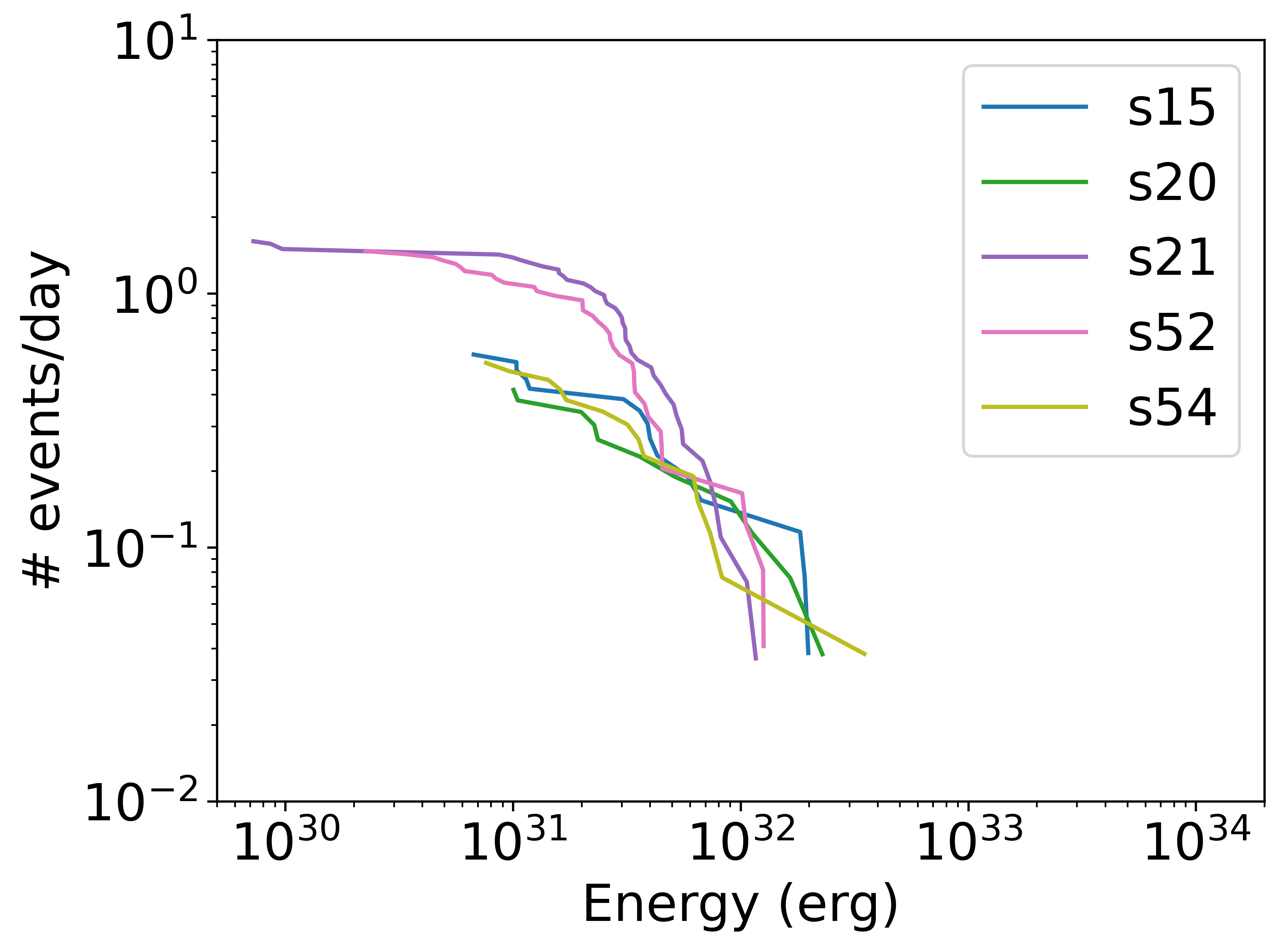}
    \caption{Cumulative events per day vs energy for non-comparable sectors of G 258-33.}
    \label{fig:flareG258ks}
\end{figure}

We present a detailed analysis of two stars with the most observational sectors and most flaring: G 227-22 (M5.0V) and G 258-33 (M4.5Ve). We selected these stars from among eight stars with more than 20 sectors, because they exhibited the highest average number of flares— approximately $68$ for G 227-22 and $38$ flares G 258-33, respectively, compared to an average of about $12$ for the others. These stars possess data coverage spanning 27 and 24 sectors, respectively, which facilitated a comprehensive analysis of their energy distribution, flare activity, and general stellar behavior over time.\\
As seen in Fig. \ref{fig:flareG227}, G 227-22 exhibited 2 to 4 flare events per day with energies above $10^{31}$ erg. The corresponding Fig. \ref{fig:amplitudeG227} highlights that most flares had $\tau$ lower than 1000 s.
The star G 258-33 (Fig. \ref{fig:flareG258}), appeared less active, with 0.6 to 2 daily flare events above $10^{31}$ erg. Figure \ref{fig:amplitudeG258} shows that the $\tau$ distribution for G 258-33 was similar to that of G 227-22, but with fewer flares exhibiting large amplitudes.\\ 
Finally, as for the whole sample, we calculated the slopes for these stars. For G 227-22 the average slope is $-0.92 \pm 0.24$, while  for G 258-33 it is $-1.01 \pm 0.08$. These average values are very similar and very close to the threshold of -1, which discriminates between stars dominated by high-energy or low-energy flares.\\
To assess whether each star behaves consistently across all sectors or exhibits anomalous behavior in certain cases, we tested the flare frequency for each pair of sectors using a Kolmogorov-Smirnov test. Figure \ref{fig:bothMatrices} shows the results as a matrix, where the black cells indicate the pairs that are incompatible with each other based on the chosen confidence level of 0.05.\\
For the star G 227-22, the anomalous sectors identified were 15, 18, 25, 60, and 73 (5 out of 27 sectors). In Fig. \ref{fig:flareG227ks}, we show the cumulative frequency energy distribution specifically for these sectors. Notably, sectors 25 and 60 are not comparable even with the other anomalous sectors, as their light curves exhibit lower sensitivity to less energetic flares.  For the remaining sectors, the difference appears to be due to the presence of high-energy flares in the tails of the curves.\\
For G 258-33, the anomalous sectors were 15, 20, 21, 52, and 54. The cumulative frequency energy distributions for these sectors are shown in Fig. \ref{fig:flareG258ks}.  Sectors 15, 20, and 54 are not comparable with the others, either due to the limited sensitivity of the light curve to low-energy flares or the presence of high-energy flares. We also investigated whether a temporal periodicity could be identified in the properties of the distributions of the two stars, which would indicate the presence of stellar cycles, but found no significant evidence of periodicity. 


\section{Discussion}\label{discussion}
\begin{figure*}
    \centering
    \subfloat[]{
        \includegraphics[width=0.45\hsize]{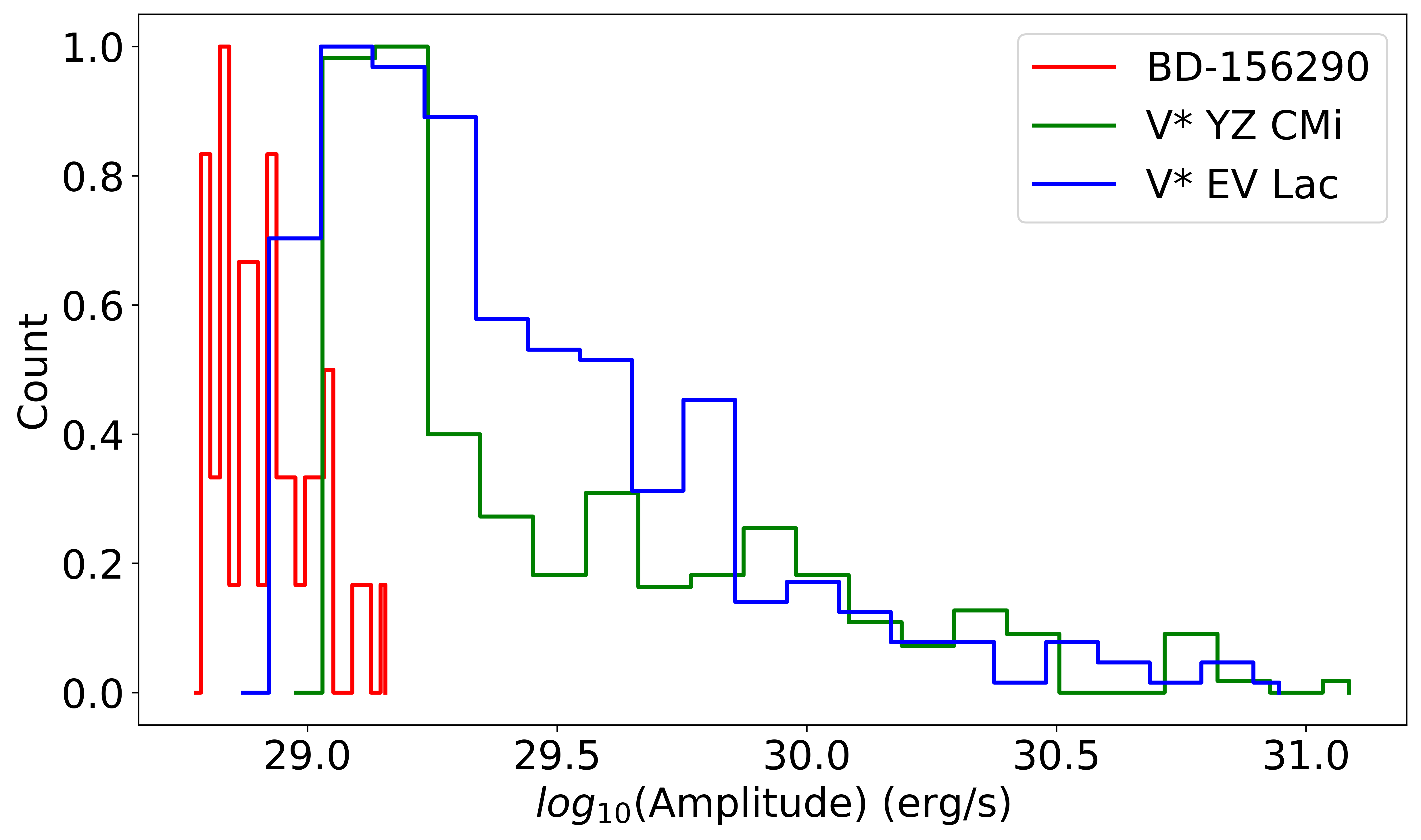}
        \label{fig:histamplitude}
    }
    \subfloat[]{
        \includegraphics[width=0.45\hsize]{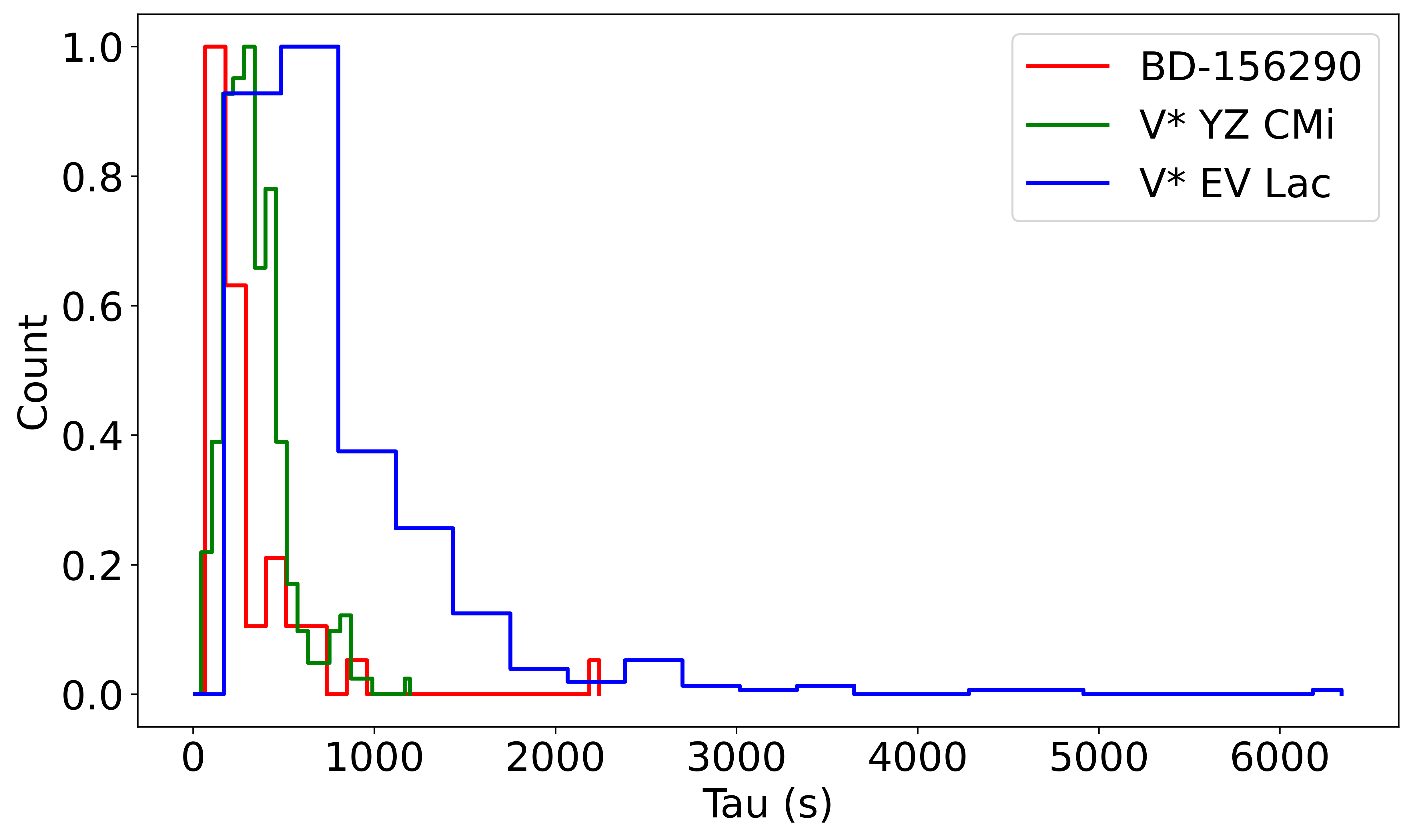}
        \label{fig:histtau}
    }
    \caption{(a) Normalized distribution of amplitude for the three example stars. (b) Normalized distribution of $\tau$ for the three example stars.}
    \label{fig:histogramamplitudetau}
\end{figure*}

\begin{figure*}
    \centering
    \subfloat[]{
        \includegraphics[width=0.45\hsize]{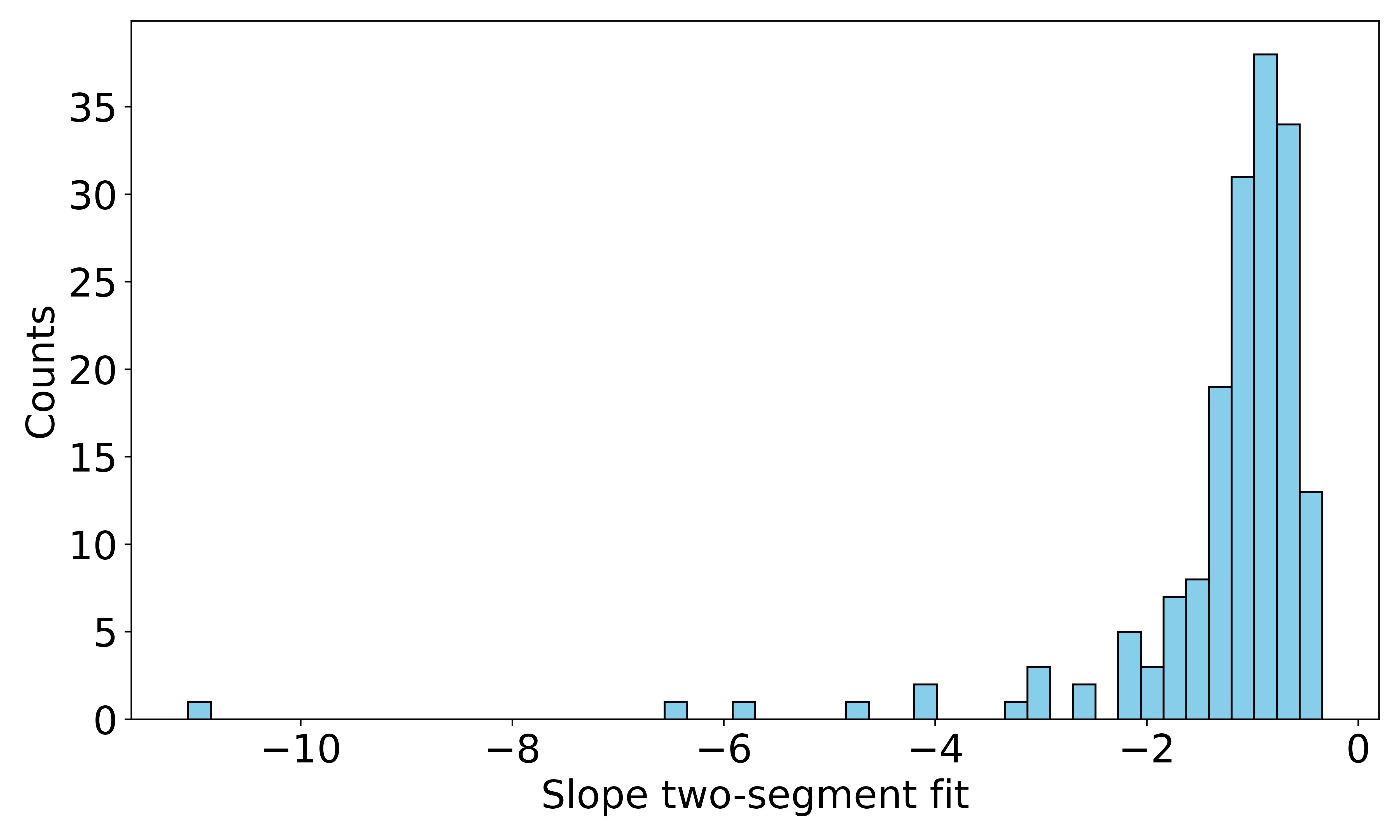}
        \label{fig:histsecond}
    }
    \subfloat[]{
        \includegraphics[width=0.45\hsize]{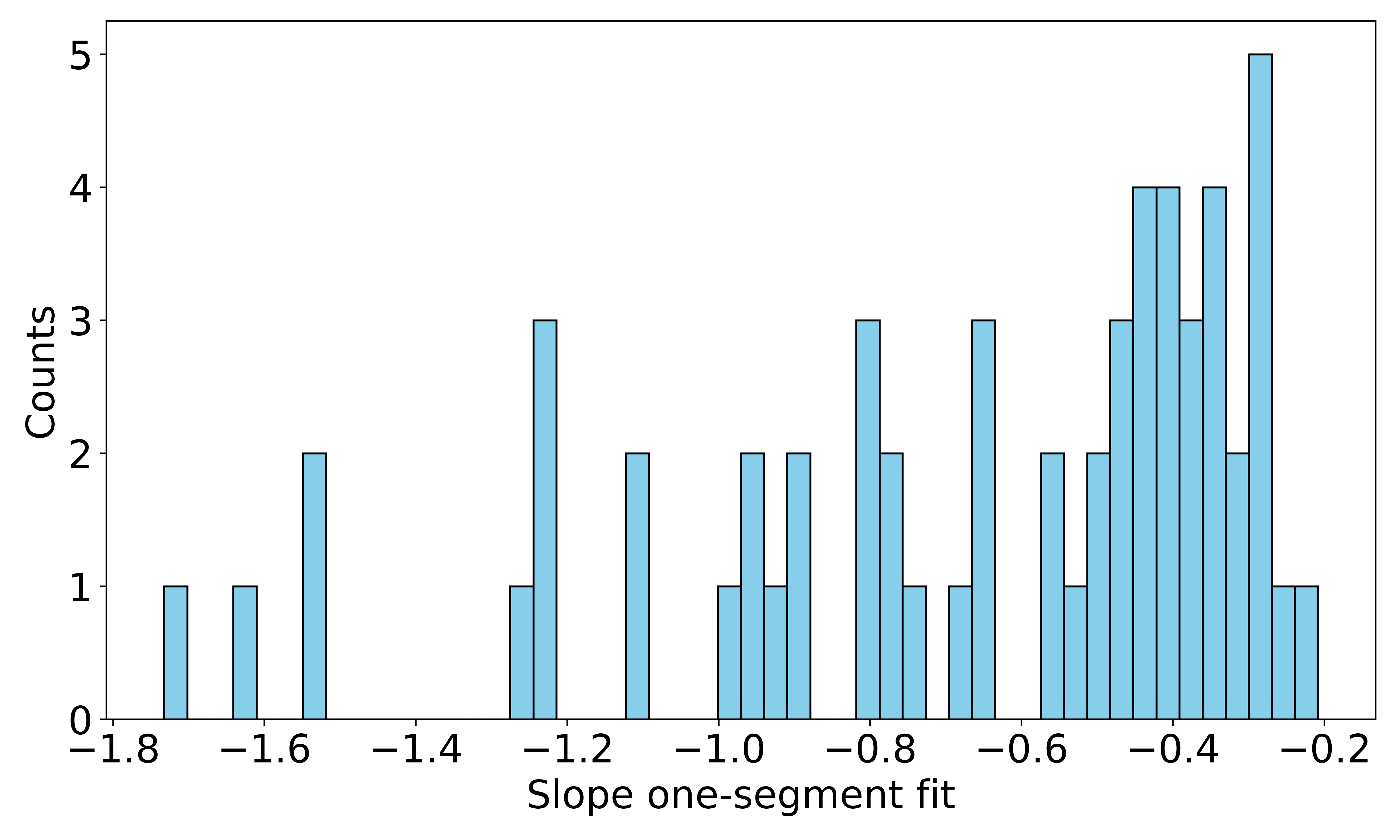}
        \label{fig:histfirst}
    }
    \caption{(a) Distribution of slopes for the two-segment model. (b) Distribution of slopes for the one-segment model.}
    \label{fig:histograms}
\end{figure*}

\begin{figure}
    \centering
    \includegraphics[width=\hsize]{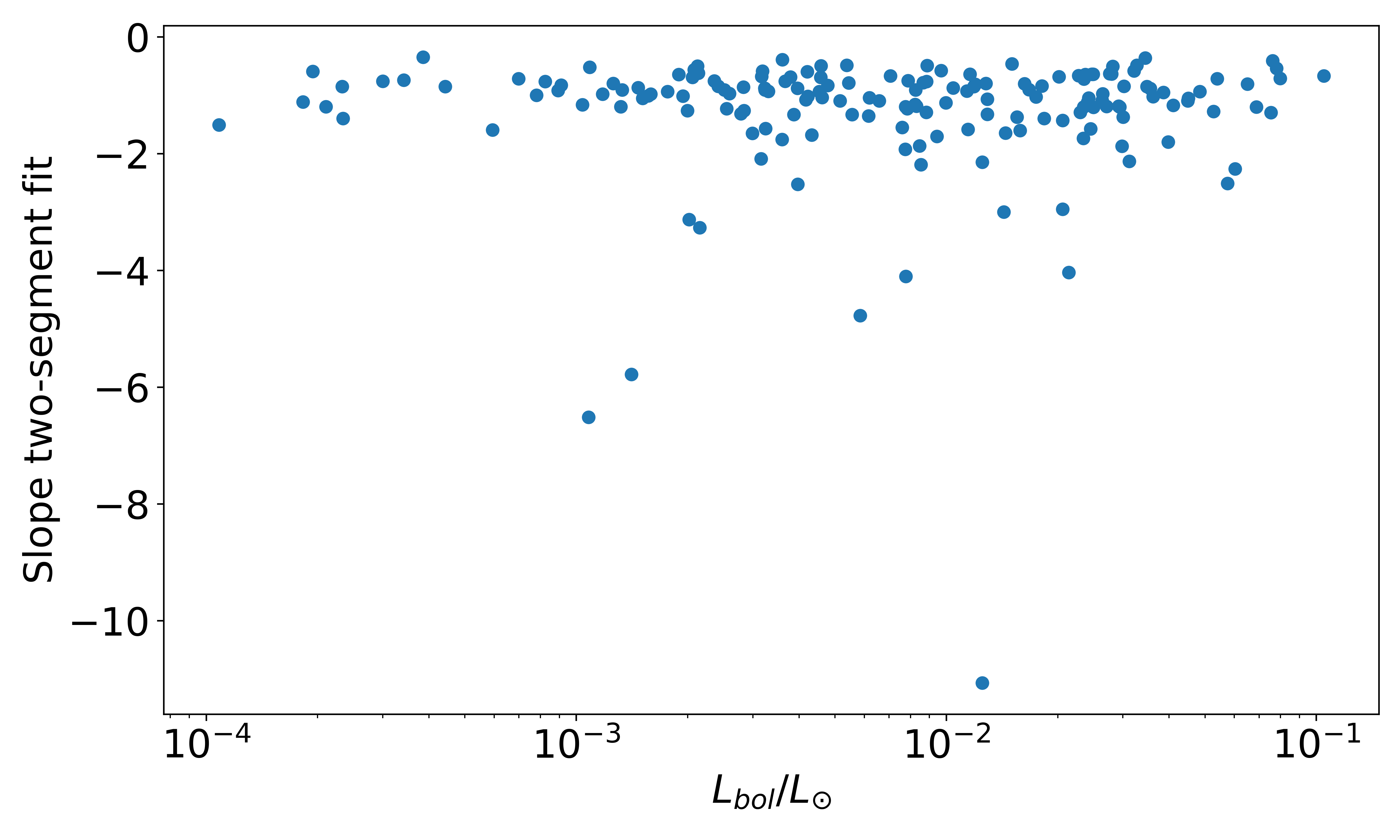}
    
    \caption{Scatter plot of slopes for the two-segment model vs $L_{bol}$. }
    \label{fig:secondmagnitude}
\end{figure}

\begin{figure}
    \centering
        \includegraphics[width=\hsize]{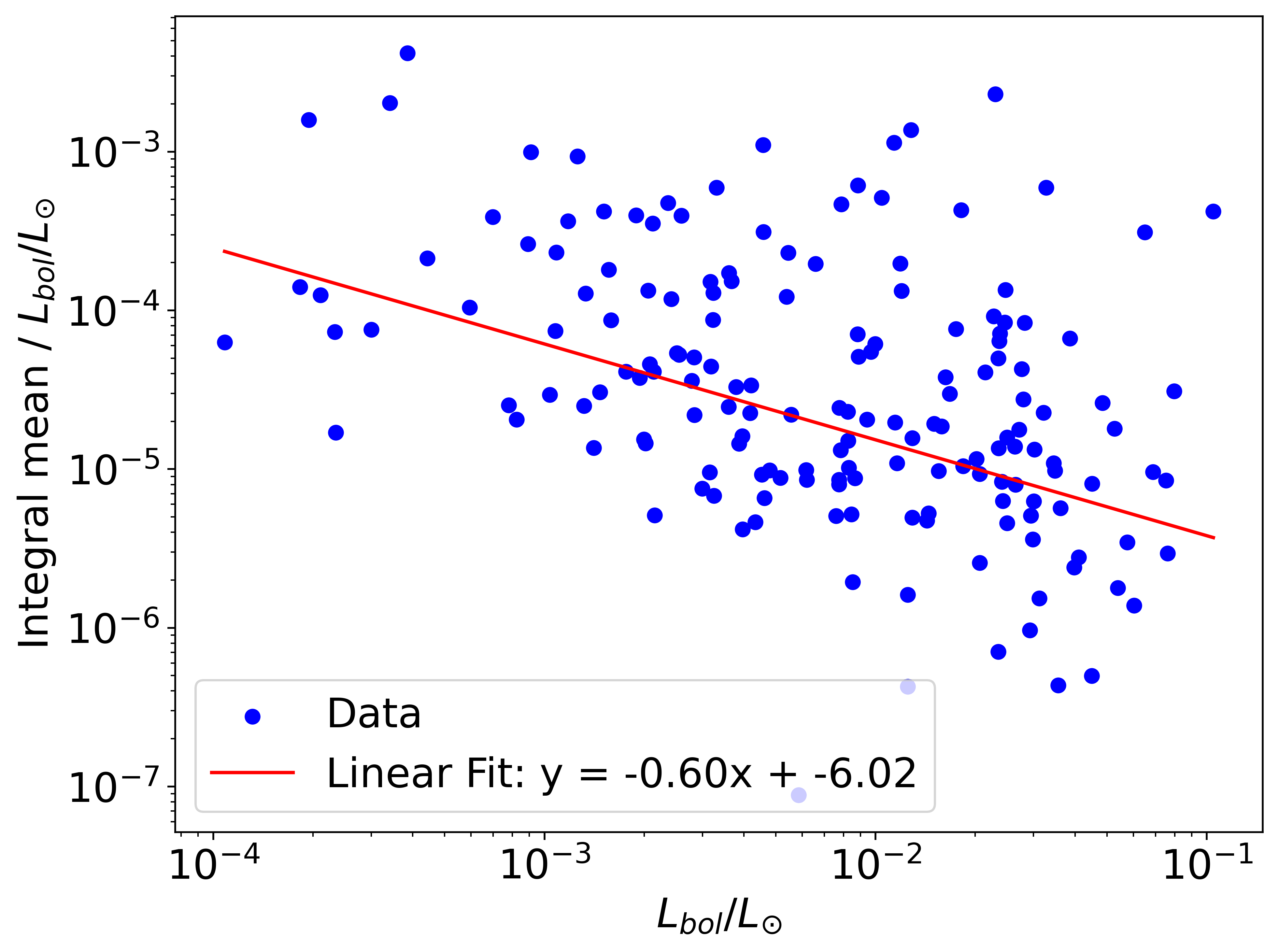}
        \caption{Integral mean of the energy/ $L_{bol}$ vs $L_{bol}$. The blue dot represent the integral mean of the energy above the energy break for each star. The red line represent the linear fit of the data.}
        \label{fig:integrallbol}

\end{figure}

To investigate the properties of the three sample stars discussed in section \ref{flarefrequency}, we constructed two histograms: one for the amplitudes as shown in Fig.\ref{fig:histamplitude} and another for the $\tau$ ($\tau_r+\tau_d$) of the flares as shown in Fig.\ref{fig:histtau}.\\
The amplitude plot of BD-156290 shows a pronounced peak, indicative of a small number of flares, all with similar amplitudes. In contrast, YZ CMi and EV Lac  display broader distributions. In the second histogram, which shows the distribution of  $\tau$, the three stars appear more homogeneous, except for EV Lac, which exhibits a pronounced tail towards longer $\tau$ values. This feature is responsible for the extended high-energy tail observed in the flare frequency distribution of EV Lac.\\ 
To assess the significance of the differences, a Kolmogorov-Smirnov test was applied to compare the histograms, finding that only the $\tau$ curves of BD-156290 and YZ CMi were statistically comparable.\\
To investigate the global properties of our sample, we examined the distribution of the slopes as reported in Figs. \ref{fig:histfirst} and \ref{fig:histsecond} that report slopes for the one- and two-segment models.
The second histogram does not show a well-defined peak, while the first displays a peak near -1, which is narrower than the average value reported in Table \ref{tab:tabslopesmean}, indicating an asymmetry in the distribution.\\
Figure \ref{fig:secondmagnitude} shows the scatter plot of slopes from the two-segment model versus $L_{bol}$.
Notably, the figure presents a larger number of data points with a broader scatter.\\
We do not observe stars dominated by low-energy flares at these faint luminosities, in contrast to the brighter luminosity regime, where such stars can be present. (See Fig. \ref{fig:secondmagnitude}).
To verify that this observed effect is not due to differences in bolometric luminosity among the stars in our sample, we calculated the integral of the flare luminosity detected beyond the energy break, normalized to the bolometric luminosity. Although this quantity is affected by sensitivity limitations, it provides an indication of each star's ability to convert available energy into energy released via flares. 
Figure \ref{fig:integrallbol}  shows this quantity as a function of the star's bolometric luminosity, suggesting that fainter stars are more efficient at releasing energy through flares. This trend is illustrated by the linear fit (red line), with individual stars shown as blue dots.\\

\begin{figure}
    \centering
    \includegraphics[width=\hsize]{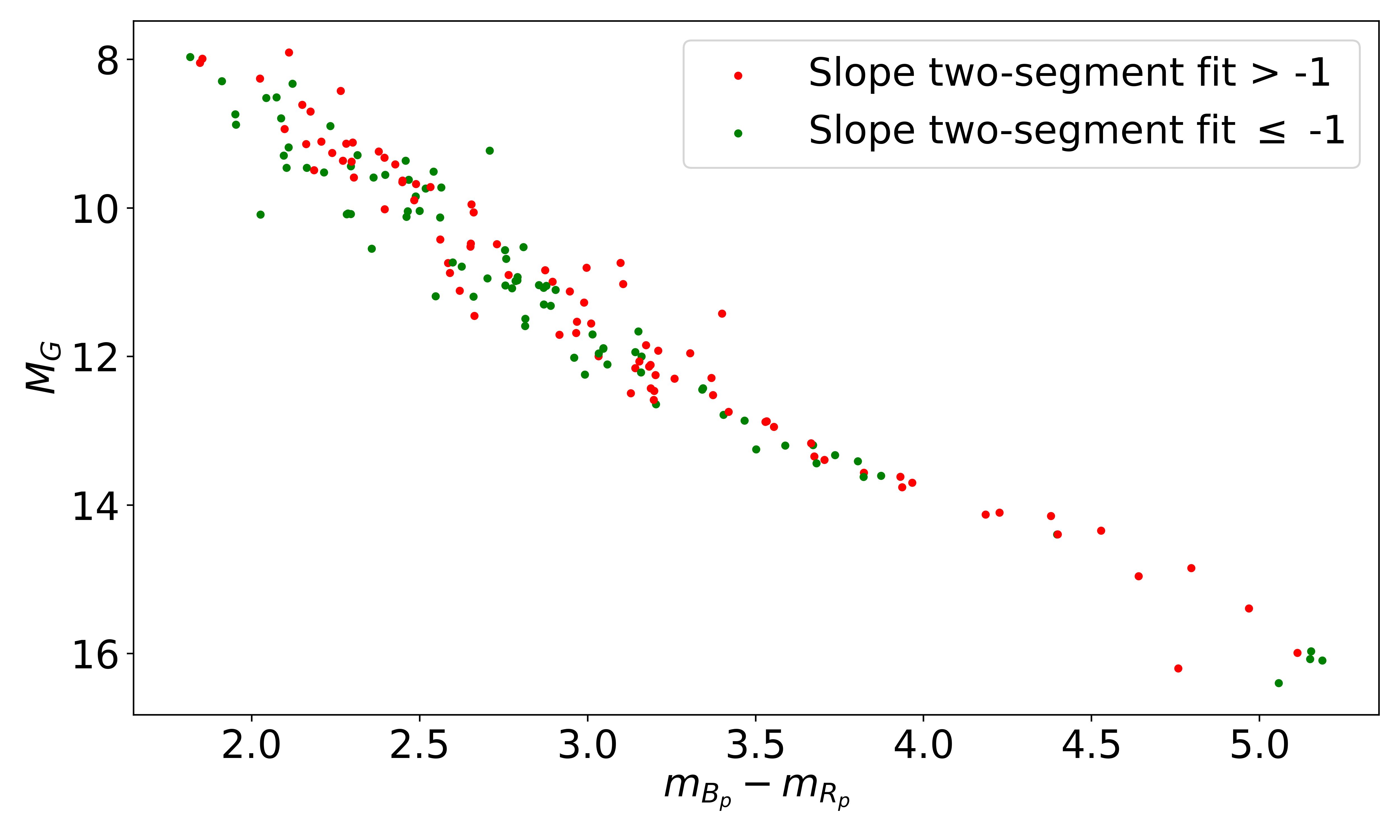}
    \caption{HR diagram of the analyzed star. The red dot are the stars with slope >-1, while the green are the stars with slope $\leq$-1.}
    \label{fig:hrslope}
\end{figure}

To further connect the flare properties with the fundamental stellar properties, Fig. \ref{fig:hrslope} presents the Hertzsprung–Russell (HR) diagram of the analyzed stars. Red dots indicate stars with a two-segment slope >-1 (steeper slopes), while the green dots represent stars with a slope $\leq$-1 (flatter slopes).\\
The analysis of Figs. \ref{fig:integrallbol} and \ref{fig:hrslope} indicates that the activity of the faint stars is dominated by a low number of high-energy flares corresponding to flatter slopes, with the exception of the coldest stars. In contrast, the activity of a substantial fraction of bright stars is dominated by a high number of low-energy flares, which are characterized by steeper slopes.  \\
To quantify the activity level of each star, we introduce the figure of merit GF.01, defined as the energy value corresponding to a flare frequency of 0.1 events per day. This value can be measured for the majority of our targets, as given by the following equation:

\begin{equation}
    GF.01= \frac{0.1-q}{m}/L_{bol},
\end{equation}

where $m$ is the slope obtained in Sect. \ref{slopecalculations}, $q$ is the intercept, $L_{bol}$ is the bolometric luminosity of the star, and 0.1 indicates the characteristic energy of flares occurring on average every 0.1 days.

\begin{figure}
    \centering
    \includegraphics[width=\hsize]{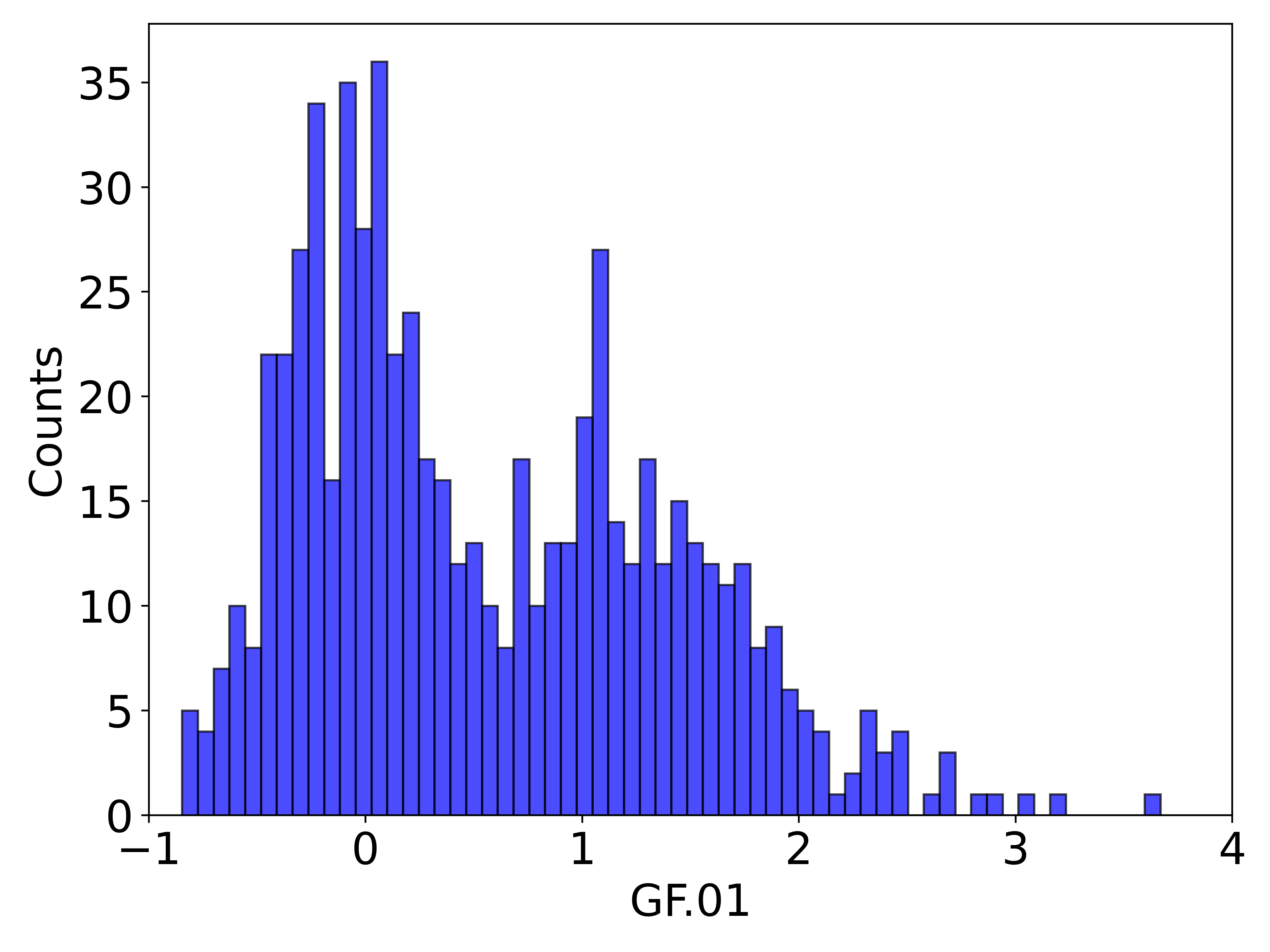}
    \caption{Distribution of GF.01. The minimum separating "quiet" and "active" stars occurs at GF.01=0.64.}
    \label{fig:histgindex}
\end{figure}

\begin{figure}
    \centering
    \includegraphics[width=\hsize]{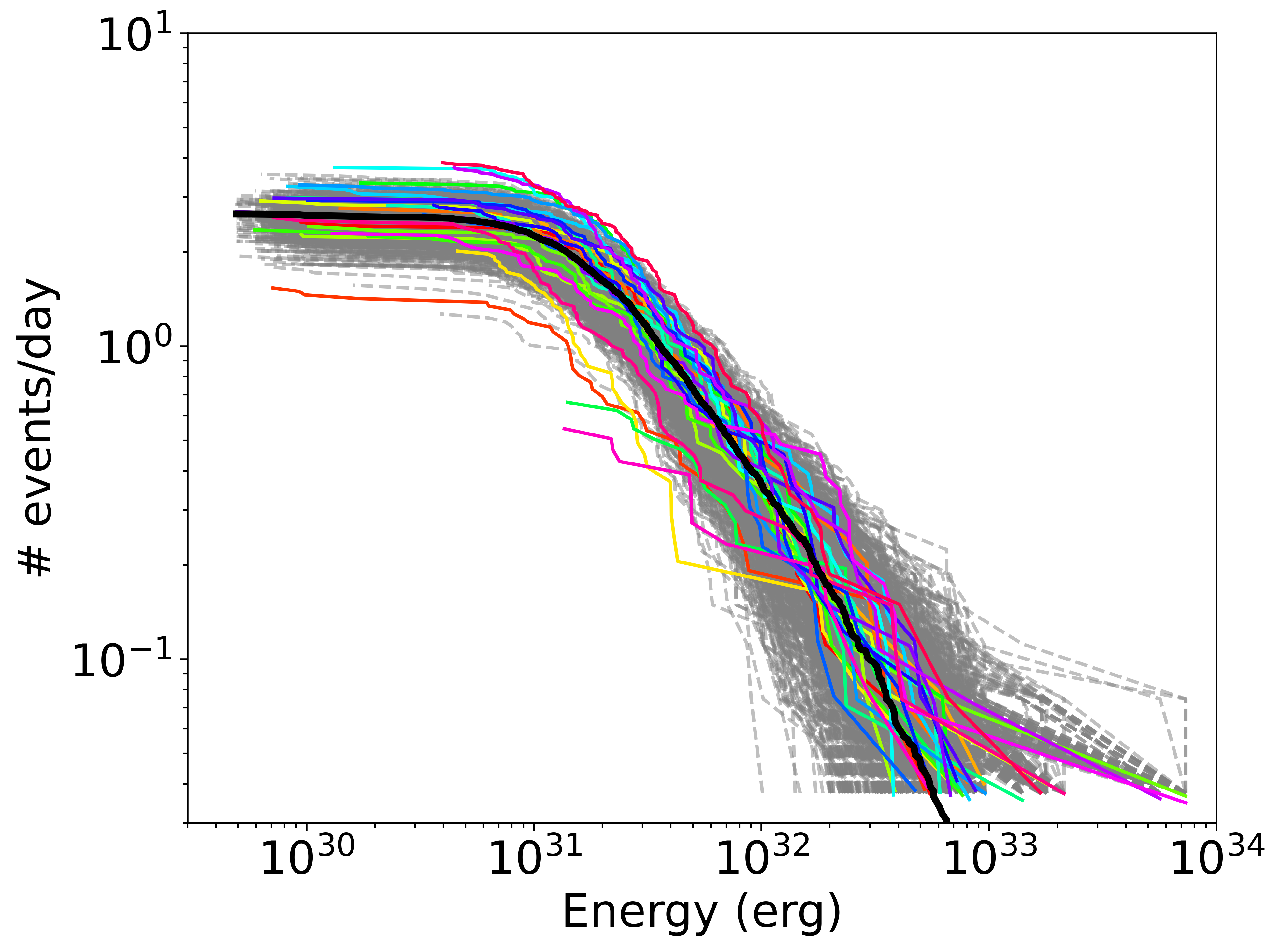}
    \caption{Cumulative events per day vs energy simulated for G 227-22. The gray line represent the simulated curve. The colored lines represent the original cumulatives and the black line represent the median cumulative.}
    \label{fig:simulated}
\end{figure}

\begin{figure}
    \centering
    \includegraphics[width=\hsize]{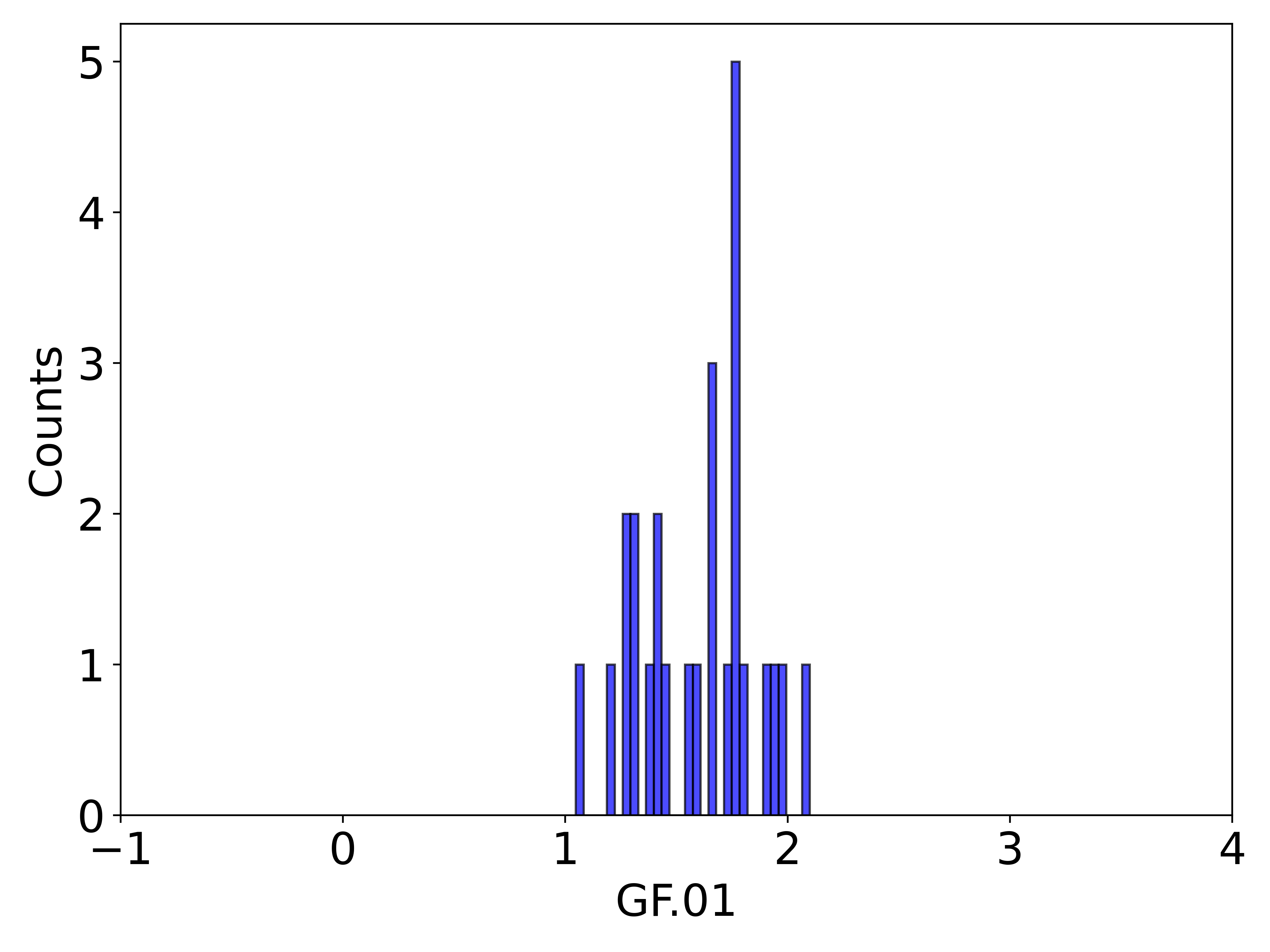}
    \caption{Distribution of GF.01 for G227-22.}
    \label{fig:histgindexg227-22}
\end{figure}

\begin{figure}
    \centering
    \includegraphics[width=\hsize]{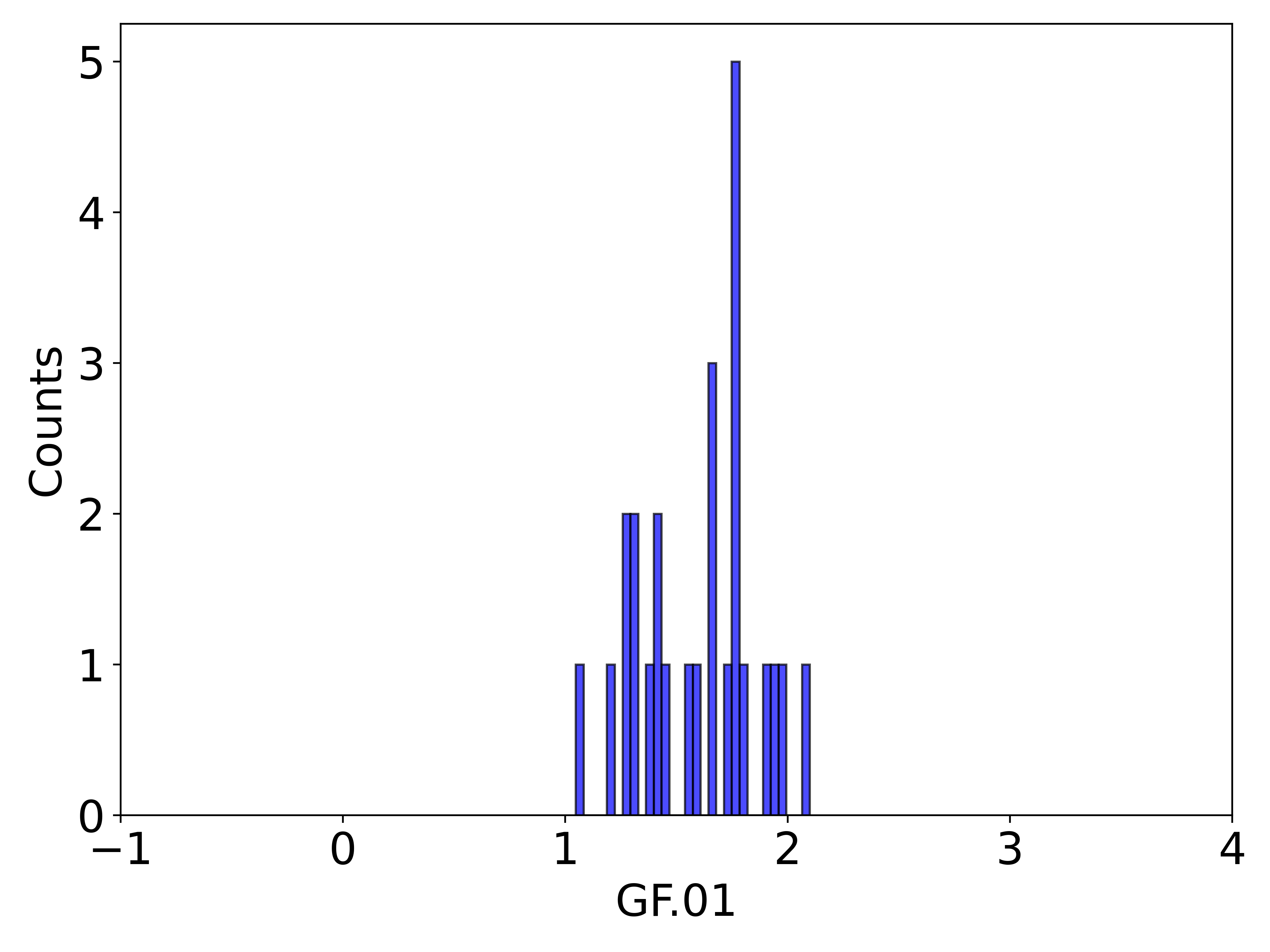}
    \caption{Distribution of GF.01 for G258-33.}
    \label{fig:histgindexg258-33}
\end{figure}

The GF.01 value serves as a flaring activity index, normalized to bolometric luminosity to enable a direct comparison of flare properties across stars with different luminosities and to remove the direct dependence on $L_{bol}$. This normalization accounts for differences in stellar luminosity, thereby enabling an assessment of how efficiently a star converts its available energy into flare energy. A higher GF.01 value indicates that, on average, the flares are more energetic, whereas a lower GF.01 value corresponds to less energetic flares.\\
Figure \ref{fig:histgindex} shows the distribution of GF.01 for sectors with two-segment slope fits, revealing a bimodal structure with a minimum of approximately 0.64. This suggests the presence of two distinct stellar populations.
Figure \ref{fig:slopevsgindex} plots GF.01 against the two-segment model slopes, showing that stars belonging to the first population tend to have steeper slopes, while those in the second population predominantly exhibit flatter slopes. The black dotted line indicates the minimum of the distribution from Fig. \ref{fig:histgindex}. Notably, the region corresponding to high GF.01 values (i.e., stars with more energetic flares) is sparsely populated at steep slopes. This region is populated by easily detectable flares, and their absence in the upper left corner of the plot suggests that the observed bimodality is not driven by selection effects or observational biases.
Fig. \ref{fig:energyvsgindex} illustrates the correlation between GF.01 and bolometric luminosity, indicating that the first population consists mainly of bright stars, whereas the second population is primarily composed of faint stars.
Together, these results suggest that faint stars exhibit flatter flare energy slopes, indicating that their flare activity is dominated by high-energy events. Conversely, bright stars tend to include both quiet and active stars. 
Figures \ref{fig:slopevsgindex} and \ref{fig:energyvsgindex} are in the appendix.\\
We analyzed in detail two stars that have been observed in multiple sectors, both exhibiting a high number of flares. Overall, the flare frequency distributions show no significant differences, indicating minimal variability on the observed timescales (around 5 years). To verify this, we constructed the global flare frequency for star G 227-22, considering all flares detected across the 27 sectors. We assume that this is the true flare frequency distribution; we generated 1000 simulated curves by randomly selecting the energy values from each sector. For each simulation, the number of events chosen was determined using a Gaussian distribution centered on the mean number of events per sector (68) with a standard deviation equal to its square root. The simulated curves, shown superimposed on the observational data in Fig.\ref{fig:simulated}, follow the observed data, suggesting that each data sector provides a representative sample of stellar activity. Curves from sectors with low sensitivity fall outside the gray area and therefore differ from the average population due to observational biases rather than intrinsic differences.  Since the star's activity shows no significant variation over time, the flare fluctuations are consistent with its characteristic behavior.\\
We also analyzed the GF.01 distributions for G 227-22 and G 258-33 in Figs. \ref{fig:histgindexg227-22} and Fig. \ref{fig:histgindexg258-33}, respectively. Both stars belong to population 2, as expected, showing the high-energy-driven flare activity typical of faint stars. 
\section{Conclusions}\label{conclusions}
In summary, we analyzed a volume-limited sample of 173 M dwarfs within 10 parsecs, covering a total of 751 TESS sectors. Using an updated methodology based on \cite{colombo2022short} and the VOSA SED analyzer for bolometric luminosity computation, we detected a total of 17229 flares. We developed a tool to determine the slopes and energy breaks of the derived cumulative curve.\\
A detailed analysis of the flare energy distributions of was conducted and we highlighted three representative cases. Subsequently, we focused on analyzing the energy breaks and slopes
for the entire sample.
For the two-segment fit, the average slope is -1.23$\pm$1.32 indicating that the short-term variability of most stars is dominated by low-energy flares. The one-segment fit yields an average slope of  -0.79$\pm$0.64, which is biased by the limited sensitivity of TESS.\\
Next, we analyzed the relationship between slopes and $L_{bol}$. We find that the activity of the faint stars is dominated by a small number of high-energy flares. In contrast, bright stars include both those dominated by a large number of low-energy flares and those with a small number of high-energy flares.\\
To characterize flare properties, we defined the Flare Energy Index (GF.01), whose bimodal distribution reveals two stellar populations: faint stars with flatter slopes dominated by high-energy flares, and bright stars including stars with steeper slopes driven by low-energy flares.
Finally, we present a detailed analysis of two highly observed and active stars, G 227-22 (M5.0V) and G 258-33 (M4.5Ve), which found no evidence of periodic time variability in their flare frequency distribution.


\section{Data Availability}
Table 3 shows the names of the stars analyzed, the TESS sectors, the number of flares, the $L_{bol}$, the slopes, and the energy breaks. The value N/A indicates the one-segment fit cases where no energy break is defined. The nan value on the slope indicates the cases with one or zero flares per sector.
Table 3 is only available in electronic form at the CDS via anonymous ftp to cdsarc.u-strasbg.fr (130.79.128.5) or via http://cdsweb.u-strasbg.fr/cgi-bin/qcat?J/A+A/.

\begin{acknowledgements}
   G. M. acknowledges support from the European Union - Next Generation EU through the grant n. 2022J7ZFRA - Exo-planetary Cloudy Atmospheres and Stellar High energy (Exo-CASH) funded by MUR - PRIN 2022 and the ASI-INAF agreement 2021-5-HH.2-2024.\\
   L. P. acknowledges support from the Italian Ministero dell'Università e della Ricerca and the European Union - Next Generation EU through project PRIN 2022 PM4JLH ``Know your little neighbours: characterising low-mass stars and planets in the Solar neighbourhood''.\\
   This work has made use of data from the European Space Agency (ESA) mission {\it Gaia} (\url{https://www.cosmos.esa.int/gaia}), processed by the {\it Gaia} Data Processing and Analysis Consortium (DPAC, \url{https://www.cosmos.esa.int/web/gaia/dpac/consortium}). Funding for the DPAC has been provided by national institutions, in particular the institutions participating in the {\it Gaia} Multilateral Agreement.\\
   This research has made use of the SIMBAD database, operated at CDS, Strasbourg, France \citep{wenger2000simbad}.\\
   This research made use of Lightkurve, a Python package for Kepler and TESS data analysis \citep{2018ascl.soft12013L}.\\
   This publication makes use of VOSA, developed under the Spanish Virtual Observatory (\url{https://svo.cab.inta-csic.es}) project funded by MCIN/AEI/10.13039/501100011033/ through grant PID2020-112949GB-I00. VOSA has been partially updated by using funding from the European Union's Horizon 2020 Research and Innovation Programme, under Grant Agreement nº 776403 (EXOPLANETS-A).
\end{acknowledgements}

%

   \bibliographystyle{aa} 
   \bibliography{bibliografia.bib} 
%
\begin{appendix}
\section{GF.01 Plot}
\begin{figure}
    \centering
    \includegraphics[width=\hsize]{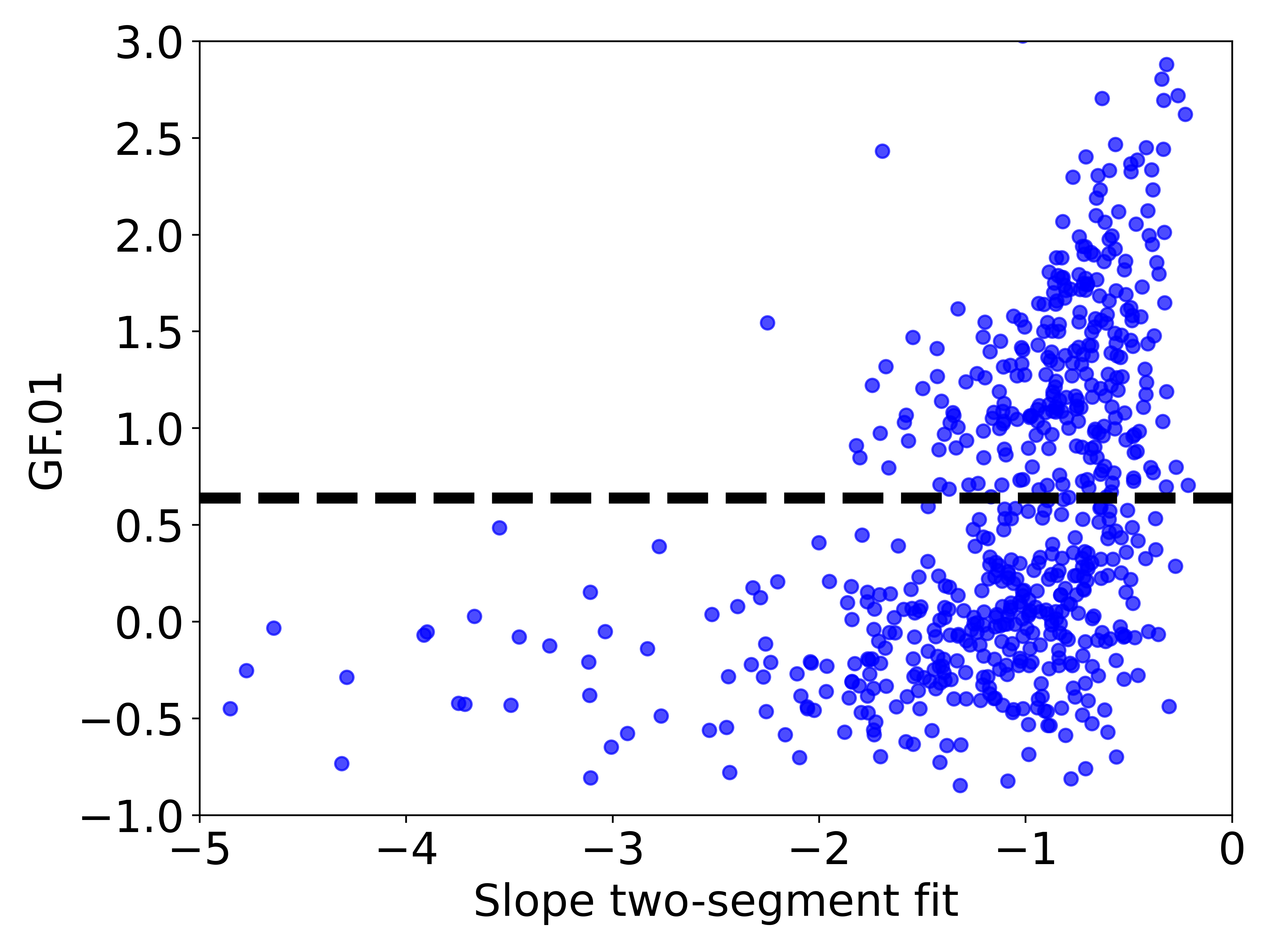}
    \caption{Scatter plot of GF.01 vs two-segment model slopes. The black dotted line represent the value of the minimum of the distribution of GF.01.}
    \label{fig:slopevsgindex}
\end{figure}

\begin{figure}
    \centering
    \includegraphics[width=\hsize]{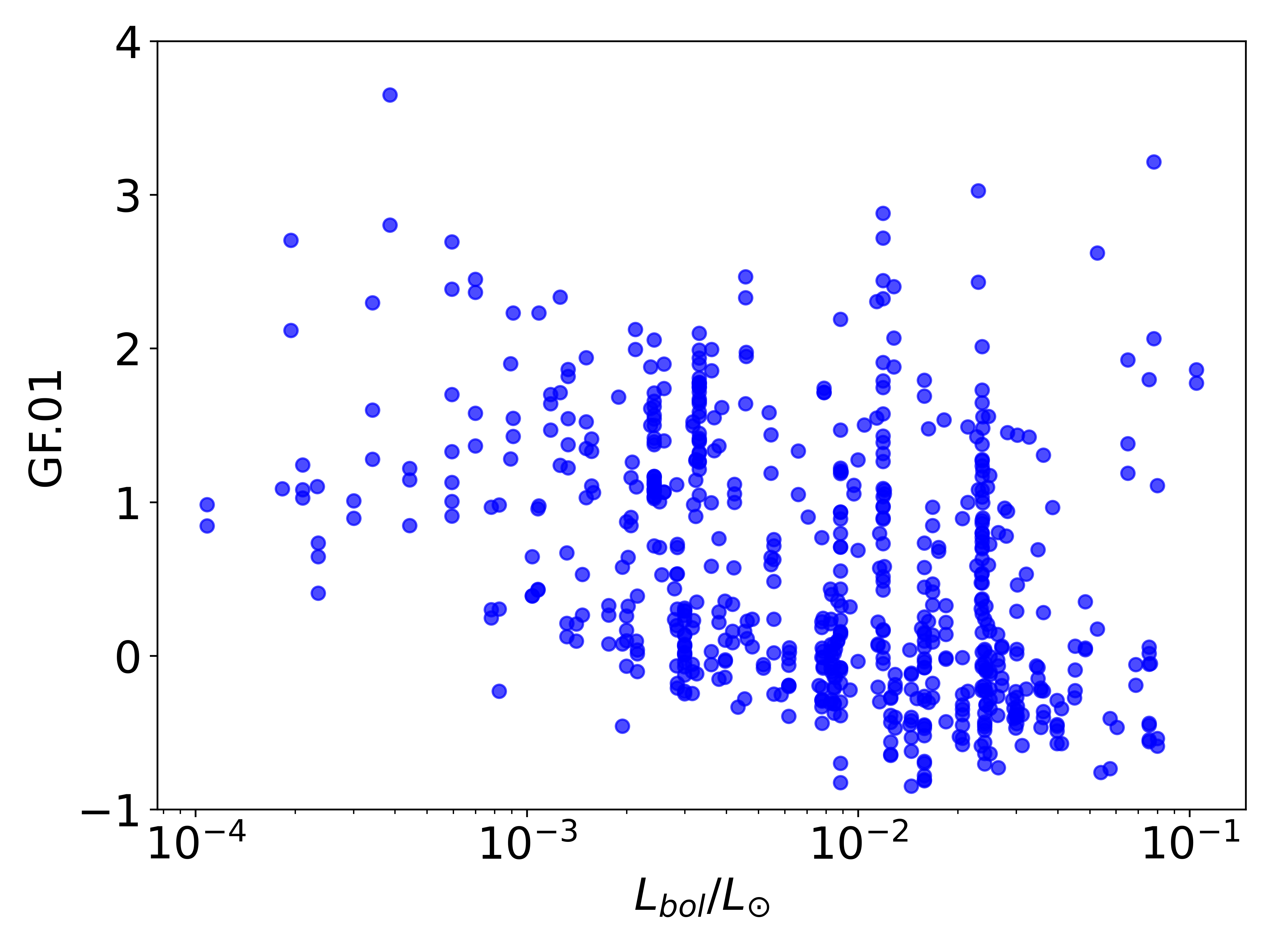}
    \caption{Scatter plot of GF.01 vs $L_{bol}$}
    \label{fig:energyvsgindex}
\end{figure}

Fig. \ref{fig:slopevsgindex} plots GF.01 vs the two-segment model slopes. The black dotted line represent the value of the minimum of the distribution from Fig. \ref{fig:histgindex}. 
Fig. \ref{fig:energyvsgindex} illustrates the correlation between GF.01 and bolometric luminosity.


\end{appendix}

\end{document}